# NUMERICAL INVESTIGATION OF FLOW STRUCTURES AROUND A CYLINDRICAL AFTERBODY UNDER SUPERSONIC CONDITION

Pratik Das, Ashoke De*
Department of Aerospace Engineering, Indian Institute of Technology, Kanpur, India-208016

*Corresponding author: Ph:+91-5122597863, Fax: +91-5122597561
E-mail: ashoke@iitk.ac.in

## ABSTRACT

Large Eddy Simulation (LES) with dynamic Smagorinsky model has been applied to numerically investigate the complicated flow structures that evolve in the near wake of a cylindrical after body aligned with a uniform Mach 2.46 flow. Mean flow field properties obtained from numerical simulations, such as axial velocity, pressure on base surface, have been compared with the experimental measurements as well as with other published results. It has been found that standard k-epsilon model fails to predict the flow properties in the recirculation region where better agreement has been observed between the data obtained from LES and experimental measurements. Flow Statistics like turbulent kinetic energy and primary Reynold's stress have also been calculated and compared with the results obtained from experiments in order to quantitatively assess the ability of LES technique to predict the turbulence properties of flow field in the highly compressible shear layer region. The data obtained from LES has been further analyzed to understand the evolution of coherent structures in the flow field. Proper Orthogonal Decomposition (POD) of the data obtained from central plane in the wake region has been performed in order to reveal the most energetic structures present in the flow field.

## NOMENCLATURE

| | |
|---|---|
| $C_p$ | Pressure co-efficient |
| $C_s$ | Smagorinsky constant |
| $R_0$ | radius of the cylindrical after-body |
| $S_{ij}$ | mean flow strain rate tensor |
| $T$ | Temperature |
| $U$ | mean velocity |
| $U_a$ | Axial component of mean Velocity |
| $k$ | Turbulent kinetic energy |
| $p$ | Pressure |
| $t$ | Physical time |
| $u_i$ | Velocity vector |
| $u_i'$ | Fluctuating component of velocity vector |
| $\Omega_{ij}$ | Rotation tensor |
| $\Delta$ | Filter size |
| $\nu$ | Kinematic viscosity |

**Super-scripts**

| | |
|---|---|
| $-$ | Averaged variable |

**Sub-scripts**

| | |
|---|---|
| $1$ | Free-stream condition |
| $axial$ | Axial component of a vector |
| $average$ | Time averaged variable |
| $rms$ | Root mean square |

## 1. INTRODUCTION

Bullets, projectiles, missiles, launch vehicles and rockets travelling at supersonic velocity experience massive pressure drag, or otherwise known as base drag, due to the low pressure region created after flow separation behind the base of these objects. Due to design constraints often these objects feature a blunt base with sharp corner at the rear end and the geometrical features of the rear end of these objects often closely resembles with the geometrical features of a cylindrical after-body axially aligned with the flow direction. As the flow detaches at the sharp base corner, it forms a low velocity, low pressure recirculation region in the near wake of the base and the pressure on the surface of the blunt base is reduced. Over the past years several active and passive techniques, such as boat-tailing, base cavity, base bleed and base burning have been developed to increase the base pressure and reduce overall drag on the bodies travelling at supersonic speed. Yet, in order to design optimal techniques for base pressure recovery it is imperative to achieve a thorough understating of the complex fluid dynamic processes that occur in the wake region. Thus the base flow has been studied both numerically and experimentally, for a long time because of the determinant role of the design of the rear end of these objects travelling at supersonic speed in their flight capabilities. Despite having the apparently simple geometry, accurate prediction of the flow field in the near wake region of a cylindrical after body has long eluded engineers and scientists; as the several key features in the flow field such as unsteadiness in the flow field, rapid expansion near the base corner, presence of a strong compressible shear layer, presence of the recompression shock system and interaction of the turbulent flow field with the recompression shock system, leads to an extremely complicated flow physics.

In an attempt to understand the intricate physics of supersonic base flows, many experimental and numerical studies have been performed over the past few decades.

Among these, most significant experimental studies have been performed by Herrin and Dutton [1]. They have performed an extensive study of mean and turbulent flow properties of the near wake region behind a cylindrical after-body of 63.5 mm diameter, in perfect axial alignment with a Mach 2.46 flow. Their experimental facility was specifically designed to maintain the axial alignment while reducing the effects of support strings on the flow field. The experiments performed by Herrin and Dutton [1] provides an excellent database for validating the results obtained from CFD (Computational Fluid Dynamics) solvers and assessing the performance of different numerical techniques and mathematical models applied to solve this complicated problem.

In the past two decades several numerical studies have also been performed mostly focused on replicating the experimental data obtained by Herrin and Dutton [1]. Earlier attempts to numerically investigate the supersonic base flow were based on RANS (Reynolds Averaged Navier Stokes) based simulations with different turbulence models. Sahu [2] performed RANS study of base flow on 2D axisymmetric geometry and found that the two equation k-epsilon turbulence model was able to perform better than algebraic models. Chuang and Chieng [3] demonstrated that among three higher-order turbulence models, i.e. the low Reynolds-number form of a standard two-equation model, the two-layer algebraic stress model, and the Reynolds-stress model; the Reynolds Stress Model performed better than the other two. Benay and Servel [4] assessed the performance of two equation k-ω turbulence model in the context of supersonic base flow. Papp and Ghia [5] applied RNG turbulence model to simulate the axisymmetric base flow. Feo & Shaw [6] used commercial solver (FLUENT) to compare performance of Spalart-Almaras, k-ω, and RSM (Reynold Stress Models). From their study they found that RSM approach can render satisfactory results. Dharavath et al. [7] performed RANS based numerical study of massively separated flow and they observed that the renormalized group k-ϵ turbulence model performs better compared to the k-ω turbulence model. From all these previous studies it has been found that, though some of the RANS based turbulence models were able to predict the mean flow properties with moderate success, standard turbulence models failed to predict the flat pressure profile on the base surface and often overestimation of velocity has been observed in the near wake region. Standard turbulence models fail to predict the expansion of compressible shear layer formed after detachment of flow at the base corner. Under prediction of turbulent production term in the shear layer leads to prediction of a shorter recirculation region and eventually a much lower pressure level at the base surface. Turbulence models with compressibility correction were successful in predicting the mean flow properties but the pressure predicted on the base surface had radial variations, due to increased centreline velocity, on contrary to the flat base pressure profile observed by Herrin and Dutton [1]. Thus, the results obtained from RANS simulations have great model dependency.

With the advent of modern high performance computers in the dawn of twenty-first century, many researchers [8-14] has performed unsteady numerical simulations with advanced numerical techniques such as LES, DES (Detached Eddy Simulation) and many other hybrid techniques which require intensive computing power. Forsythe et al. [8] performed DES simulation with compressibility corrections of the base flow with similar geometry and boundary conditions. Kawai and Fuiji [9-10] did a comparative study on performance of LES, MILES, RANS/LES hybrid methods. They considered two different values for Smagorinsky constant ($C_s$), 0.12 and 0.24 and had presented the conclusion that the higher value of Smagorinsky constant is optimal for compressible flow than incompressible flows. Simon et al. [11] compared the performance of LES, DES, and hybrid RANS/LES for the supersonic base flow case and discussed the effect of different numerical parameters relevant to hybrid methods on the results. Rodebaugh et al. [12]

performed DDES calculations based on an extended k-ε RANS model to simulate various aero-propulsive flows and obtained acceptable match with the experimental data. From these afore-mentioned studies it is evident that advanced unsteady numerical simulations like LES , DES and other hybrid methods are able to predict the mean flow properties with reasonable accuracy while successfully predicting the shear layer thickness at different locations of the near wake region and the flat pressure profile on the base surface. More recently, Luo et al. [14] deployed relatively less computationally expensive 3-D PANS (partially averaged Navier-stokes) models based on the Menter-SST turbulence model and the Wilcox k-ω model to simulate supersonic base flow and were able to obtain satisfactory results, however their study showed that the results were dependent on the resolution control parameter and increase in resolution or decrease in the value of the resolution control parameter did not always ensure improvement in the match between the computed results and the experimental data.

All of these previous studies have been primarily focused on how accurately these numerical methods are able to predict the mean flow-field in the near wake region of the cylindrical after body whereas a little effort has been made to identify the flow structures present in the turbulent wake region. Sandberg and Fasel [15] performed a DNS study of transitional supersonic base flow at a Mach no. of 2.46 where they simulated only half of the cylinder and kept the Reynolds number around $10^5$ due to limitation in computational resources. In their study, they have investigated the effects of coherent structures associated with different azimuthal modes on the mean flow, in particular on the base pressure. It is quite evident from these available literatures that the most of these previous numerical works were mainly directed towards the investigation of mean flow field successfully, while a little attention were given to the evolution of flow structures in the wake region of the cylindrical after body placed in supersonic flow. Study of the flow structures in supersonic regime still remains a sparsely explored area of research. Hence, in the current work, the evolution of vortical structures in the wake of cylindrical after-body placed in a supersonic flow has been investigated; while LES is carried out in order to numerically resolve the large scale flow structures present in the wake region. We have chosen the experiment performed by Herrin and Dutton [1] as our test case for validation. Performance of RANS (*k*-*ε* two equations model) and LES have been assessed by comparing the results obtained from simulations with experimental data. Results obtained from the LES have been further analysed to study the turbulent properties of the flow field. After-wards, in order to find out the most energetic flow structures in the flow field, energy and enstropy based POD have been performed along the central plane of turbulent flow field in the near wake region.

## 2. MATHEMATICAL MODELING:

### 2.1 Flow modeling using RANS (Two equations *k*-*ε* turbulence model ):

Chien's [16] two equations *k*-*ε* turbulence model is used in the RANS calculations. In addition to the Favre averaged governing equations for the conservation of mass, momentum, energy and species transport, two transport equations for turbulent kinetic energy (k) and turbulent dissipation rate (ε) given by the equations [1] and [2] are solved to get the local values of k and ε.

$$\rho\frac{Dk}{Dt}=\frac{\partial}{\partial x_j}\left[\left(\frac{\mu_t}{\sigma_k}+\mu\right)\frac{\partial k}{\partial x_j}\right]+\mu_t\frac{\partial u_i}{\partial x_j}\left(\frac{\partial u_i}{\partial x_j}+\frac{\partial u_j}{\partial x_i}\right)-\rho\varepsilon-2\mu\frac{k^2}{y_n^2} \qquad [1]$$

$$\rho\frac{D\varepsilon}{Dt}=\frac{\partial}{\partial x_j}\left[\left(\frac{\mu_t}{\sigma_k}+\mu\right)\frac{\partial \varepsilon}{\partial x_j}\right]+C_1\mu_t\frac{\varepsilon}{k}\frac{\partial u_i}{\partial x_j}\left(\frac{\partial u_i}{\partial x_j}+\frac{\partial u_j}{\partial x_i}\right)-C_2\rho\frac{\varepsilon^2}{k}-2\mu\frac{\varepsilon}{y_n^2}e^{-\frac{y^+}{2}} \qquad [2]$$

From the local values of k and ε, the local turbulent eddy viscosity ($\mu_t$) is calculated using the following equation,

$$\mu_t = C_\mu \rho \left(\frac{k^2}{\varepsilon}\right) \quad [3]$$

where, $C_1$, $C_2$ and $C_\mu$ are model constants.

### 2.2 Flow modeling using LES

In order to model the turbulence, LES is used where the large scale structures are resolved and the small scale structures are modeled. Hence, the Favre filtered governing equations for the conservation of mass, momentum, energy and species transport are solved in the present work [17-23]. Dynamic Smagorinsky model is used for sub-grid stress modeling [24-26], where the gradient approximation is invoked to relate the unresolved stresses to resolved velocity field and given as:

$$\widetilde{u_i u_j} - \tilde{u}_i \tilde{u}_j = -2\nu_t \bar{S}_{ij} \quad [4]$$

$$\text{Where } \nu_t = C_s^2 (\Delta)^2 |\bar{S}| \quad [5]$$

$$\bar{S}_{ij} = \left(\frac{\partial \bar{u}_i}{\partial x_j} + \frac{\partial \bar{u}_j}{\partial x_i}\right) \quad [6]$$

$$|\bar{S}| = \sqrt{2\bar{S}_{ik}\bar{S}_{ik}} \quad [7]$$

And S is the mean rate of strain. The coefficient $C_s$ is evaluated dynamically [24-26]. More details regarding the governing equations can be found in literatures [17-22].

### 2.2 Numerical Scheme

A density based, fully coupled FVM based solver has been used to solve the governing equations. A second order Low Diffusion Flux Splitting Scheme has been used to discretize the convective terms (Edwards [27]). All other spacial terms (i.e. diffusion terms) in the governing equations are discretized using second order central difference scheme, while the second order implicit temporal discretization is used. Moreover, the Low Mach number preconditioning (Weiss and Smith [278]) is used in order to effectively capture the different flow regime in the domain. The parallel processing is done using Message Passing Interface (MPI) technique. More details regarding the flow solver can be found in literatures [17-21].

### 2.3 Mathematical formulation used for POD analysis:

Objective of POD technique is to find a set of orthonormal basis vectors of an ensemble of data in a lower dimension space such that every member of the ensemble can be decomposed relative to those orthonormal bases while minimizing the error between the ensemble and its projection on the new lower dimension space. POD technique serves two purposes, firstly, it performs reduction by projecting the higher dimensional data into a lower dimensional space and secondly, it reveals the most predominant structures in the data which are often hidden in the ensemble of the data. Use of POD technique in the context of turbulent flows was introduced by Lumley [29]. The present analysis is carried out using the computationally less expensive "method of snapshots" proposed by Sirovich [30].

Let us consider N number of data sets corresponding to $N$ snapshots in time. Now $\boldsymbol{U}(\boldsymbol{x},t)$ be a fluctuating vector field with three components $U, V, W$. $\boldsymbol{U}(\boldsymbol{x},t)$ has a finite value everywhere in the domain and it is square integrable over the domain of interest. Assuming the ensemble is sufficiently large, the auto correlation tensor $R(\boldsymbol{x},\boldsymbol{x}')$ can be approximated as the following

$$R(\boldsymbol{x},\boldsymbol{x}') = \frac{1}{N}\sum_{n=1}^{N} \boldsymbol{U}(\boldsymbol{x},t^n)\boldsymbol{U}^{\boldsymbol{T}}(\boldsymbol{x}',t^n) \qquad [8]$$

Let us assume the basis mode can be written in terms of original data set as given below,

$$\emptyset(\boldsymbol{x}) = \sum_{n=1}^{N} A(t^n)\,\boldsymbol{U}(\boldsymbol{x},t^n) \qquad [9]$$

Using $R(\boldsymbol{x},\boldsymbol{x}')$& $\emptyset(\boldsymbol{x})$ , the eigen value problem can be written as

$$\sum_{n=1}^{N}\left(\frac{1}{N}\int_{\Omega}\boldsymbol{U}^{T}(\boldsymbol{x}',t^{n})\boldsymbol{U}(\boldsymbol{x}',t^{n})d\boldsymbol{x}'\right)A(t^{n})=\lambda A(t^{n}) \qquad [10]$$

Now let us define,

$$\boldsymbol{C}=C(i,j)=\frac{1}{N}\boldsymbol{U}^{\boldsymbol{T}}(\boldsymbol{x},t^{i})\boldsymbol{U}(\boldsymbol{x},t^{j})\quad i,j=1,2,3,\dots\dots\dots,N \qquad [11]$$

$$\boldsymbol{A}=A^{n}=A(t^{n})\quad n=1,2,3,\dots\dots\dots,N \qquad [12]$$

Now, equation [10] can be written as,

$$\boldsymbol{CA}=\boldsymbol{\lambda A} \qquad [13]$$

By solving the Eigen value problem presented by equation [13] $N$ mutually orthogonal Eigen Vectors $A_i, i=1,2,3,\dots\dots\dots,N$ are obtained

Normalized POD modes are given by,

$$\emptyset_{\boldsymbol{i}}(\boldsymbol{x})=\frac{\sum_{n=1}^{N}A_{i}(t^{n})\boldsymbol{U}(\boldsymbol{x},t^{n})}{\|\sum_{n=1}^{N}A_{i}(t^{n})\boldsymbol{U}(\boldsymbol{x},t^{n})\|} \qquad [14]$$

In the present study enstropy and energy based POD analyses of a 2D plane along the centerline have been performed. For enstropy based POD, the fluctuating components of vorticity vector was considered as the components of the auto correlation matrix.

$$\boldsymbol{U}=[\omega_x' \quad \omega_y' \quad \omega_z']^{T} \qquad [15]$$

And the inner product of the field is given by the following equation,

$$\left(\boldsymbol{U}(\boldsymbol{x},t^{i}),\boldsymbol{U}(\boldsymbol{x},t^{j})\right)=\int_{\Omega}\{\omega_x'(\boldsymbol{x},t^{i})\cdot\omega_x'(\boldsymbol{x},t^{j})+\omega_y'(\boldsymbol{x},t^{i})\cdot\omega_y'(\boldsymbol{x},t^{j})+\omega_z'(\boldsymbol{x},t^{i})\cdot\omega_z'(\boldsymbol{x},t^{j})\}d\boldsymbol{x} \qquad [16]$$

For compressible flow in order to perform energy based POD, it is imperative that the temperature field is also being taken into account along with the velocity components while forming the auto correlation matrix as in compressible flows, the internal energy field has significant contributions in the total energy field and considerable variation of internal energy is often observed all over the flow domain which is being considered for the POD analysis. So the fluctuation of temperature has also been taken into account along with the fluctuating velocity components.

$$\boldsymbol{U} = [u' \quad v' \quad w' \quad T']^T \tag{16}$$

While the inner product of the field is defined by the following equation,

$$\left(\boldsymbol{U}(\boldsymbol{x},t^i),\boldsymbol{U}(\boldsymbol{x},t^j)\right) = \int_{\Omega} \{u'(\boldsymbol{x},t^i)\cdot u'(\boldsymbol{x},t^j) + v'(\boldsymbol{x},t^i)\cdot v'(\boldsymbol{x},t^j) + w'(\boldsymbol{x},t^i)\cdot w'(\boldsymbol{x},t^j) + \gamma\cdot T'(\boldsymbol{x},t^i)\cdot T'(\boldsymbol{x},t^j)\}d\boldsymbol{x} \tag{17}$$

Here $\gamma$ is a scaling factor used to balance the velocity and temperature fluctuation energies. The optimal value of γ,as suggested by Lumley and Poje [31] and Qamar and Sanghi [32],is given by the following equation.

$$\gamma = \frac{\overline{\int_{\Omega}\{u'(\boldsymbol{x},t^{i})\cdot u'(\boldsymbol{x},t^{i}) + v'(\boldsymbol{x},t^{i})\cdot v'(\boldsymbol{x},t^{i}) + w'(\boldsymbol{x},t^{i})\cdot w'(\boldsymbol{x},t^{i})\}d\boldsymbol{x}}}{\overline{\int_{\Omega}\{T'(\boldsymbol{x},t^{i})\cdot T'(\boldsymbol{x},t^{i})\}d\boldsymbol{x}}} \tag{18}$$

Where ‘ $\overline{\ }$ ‘denotes averaging over time.

## 2.4 Computation Domain and Boundary Conditions

For the current study we have used two different meshes one with approximately 1.43million grid points and the other grid has 1.85 million grid points. The mesh with 1.85 million grid points has more has more axial refinement in the near wake than the mesh with 1.43 million grid ponts, thus resulting a higher resolution in the near wake than the mesh with 1.43 million grid points. Since we are using a structured multi-block solver, O-H topology was employed in the region behind the based in order to reduce skewness and achieve a better quality of mesh so that convergence issues can be dealt with even at a higher CFL number. Figure 1 illustrates dimension of the computation domain. We have considered an approach length of $4R_0$, whereas $20R_0$ is considered to be a suitable length for the domain behind the base. The outer boundary remains at a radial distance of $10R_0$ from the centreline or the axis of the base. Since the region of interest is the near wake of the cylindrical after body, fine mesh has been used near the base and walls. Further downstream and towards radially outside the domain the mesh has been stretched to optimize computation cost.

The finer mesh with 1.85 million grid points has been designed with more points in the near wake of the base in axial direction so that the dynamics of flow structures in the shear layer bounded recirculation can be resolved properly whereas the resolution in the radial direction and azimuthal direction remains same as that of the 1.43 million grid points. In Figure 2(a), it can be easily observed that the distribution of base pressure co-efficient (obtained using equation [20]) obtained from both of the meshes match closely. From Figure 2(b), it can be discerned that the log-log plot of resolved energy spectrum obtained from the mesh with 1.85 million grid points seem to approach -5/3 slope in the inertial sub range and thus the grid can be considered to be fine enough to resolve the large eddy structures in the wake region. Moreover, using the method proposed by Celik et al. [33] which uses eddy viscosity ratio as an indicator for examining the grid resolution. The formulation of this method is given by

$$LES_IQ = \frac{1}{1 + 0.05\left(\frac{\upsilon_{t,eff}}{\upsilon}\right)^{0.53}} \qquad [19]$$

In this formulation $\upsilon_{t,eff}$ denotes the effective viscosity, which is the sum of laminar and turbulent viscosity, and $\upsilon$ denotes the laminar viscosity. The LES quality index should be greater than 0.8 for good LES predictions [34]. Figure 3 depicts the LES quality index distribution throughout the domain and it confirms the quality of the present grid. Since the finer mesh has better resolution in the wake region, it is chosen for the rest of the calculations. Hence forth, we have reported the detailed results using the fine mesh (1.85 million grid points) only.

Dimensions of the base geometry and boundary conditions have been chosen according to the supersonic base flow experiments performed by Herrin and Dutton [1]. Radius of the base is 31.75mm and Mach number of the flow is 2.46 resulting a flow with Reynolds number of $2.858\times10^{6}$.

Boundary conditions applied to the domain are as follows:

| | |
|---|---|
| Pressure at Inlet | : 32078.5 Pa |
| Temperature at Inlet | : 133.012K |
| Velocity at Inlet | : 567.0m/s |

Adiabatic wall with no slip boundary condition has been applied on the cylindrical after-body. Supersonic free-stream boundary condition with pressure, velocity and temperature same as the inlet has been applied on the outer boundary of the cylindrical computation domain, while at the outlet a non-reflecting convective outflow condition is used (Akselvoll and Moin [34]). For LES simulation, the physical time step or $\Delta t$ is kept $1\times10^{-6}$ corresponds to highest value of CFL number (in the region of refinement) in order of 0.6 whereas the value of CFL drops rapidly as we move further from the base surface . Simulations are carried out for several flow-through times, while time averaging of flow field is achieved over ~30+flow-through times to obtain better statistics.

## 3. RESULTS & DISCUSSION:

### 3.1 Salient Features of the Flow Field

Though the geometry might appear to be simple, supersonic flow past a cylindrical after body possesses critical flow features like presence of expansion waves, shocks, compressible shear layer, and presence of adverse pressure gradient in the shear layer due to the recompression region and interaction of unsteady vortices with reattachment shock system. Figure 4 illustrates the salient features of the flow field in near wake of the cylindrical after body axially aligned with a supersonic flow of Mach number 2.46. In Figure 4, it is discernable that the flow turns as Prandtl-Mayer expansion waves which are formed at the base corner followed by a strong shear layer, while further downstream the flow realigns itself with the axis after passing through a recompression region. A large recirculation bubble is formed immediately downstream of the base. The turbulent boundary layer separates from the body at the base corner and forms the compressible shear layer. The low velocity (subsonic) fluid in the recirculation region is separated from the supersonic flow by the free shear layer. Further downstream the fluid within the recirculation region encounters strong adverse pressure gradient and thus, is sent back from the stagnation point towards the base. The free shear layer is characterized by intensive turbulent mixing. The dynamics of energy transfer from the supersonic flow outside the recirculation region to the low velocity flow inside the recirculation region is governed by the shear layer and thus flow properties in the wake region are greatly influenced by the shear layer. Due to high shear stresses in the free shear layer, the turbulent production term attains a high value in the shear layer thus facilitating intensive turbulent mixing and energy transfer from the supersonic flow outside the recirculation region to the subsonic flow inside the recirculation region. Thus the low velocity fluid in the recirculation region is entrained and accelerated by the shear layer. Fluid in recirculation region, closer to the shear layer, has higher kinetic energy due to the turbulent mixing and penetrates through the adverse pressure gradient in the reattachment region. But fluid in the recirculation region, away from the shear layer, has less kinetic energy and fails to penetrate through the reattachment zone and is sent back towards the base after reaching the stagnation point. As a result a low velocity and low pressure recirculation zone is formed near the base.

### 3.2 Mean Pressure and Velocity Profiles

Time averaged results obtained from LES, RANS (standard k-epsilon model) and results obtained from experiments (Herrin and Dutton[1]) have been compared and thoroughly discussed in this section. We have also compared our results with the results obtained by Simon et al. [11] in order to make a comparative assessment of our results. From application point of view, pressure at the base surface is the most crucial parameter in this study, as the objective of the study is to understand how the flow physics is governing the pressure at the base. A dimensionless pressure coefficient $C_p$ has been calculated from pressure at the base of the cylindrical after-body. Relation between base pressure and the pressure coefficient is given by Eq. 25. From Figure 5, the plot of mean axial velocity along centerline , it can be observed that the current LES study has been able to render a better prediction in the near wake region, though the length of the recirculation region has been slightly over predicted in comparison with the LES study performed by Simon et al. [11]. On the other hand, standard k-epsilon model completely fails to predict the length and location of the recirculation region. From Figure 6 it is discernable that the current study has predicted the flat pressure profile across the base surface with fair accuracy on contrary to the prediction obtained from the standard k-epsilon model which not only exhibits unacceptable amount of under-prediction of pressure but also variation of pressure across base surface along radial direction.

$$C_p = \frac{2\left[\left(\frac{P_{base}}{P_1}\right) - 1\right]}{\gamma M_1^2} \qquad [20]$$

Further investigation has been carried out by plotting the radial variation of time averaged axial velocity. From Figure 7, it can be observed that LES has been able to predict the location and thickness of shear layer with acceptable accuracy in all four locations downstream of the base. Figure 7(a) shows the axial velocity profile in radial direction at point just after the flow has separated. The shear layer thickness at this location, as it can be noticed, is very less and LES has been able to capture the axial velocity profile with fair accuracy whereas the standard *k-ε* model exhibits over prediction in axial velocity profiles and the shear-layer thickness. At the locations further downstream, the shear-layer expands and the expansion has been well predicted by LES; though at some location over prediction of mean axial velocity and little under prediction of shear layer thickness can be observed. On the other hand, RANS has completely failed to predict the shear layer expansion in the wake region, possibly because of over estimation of turbulent eddy viscosity leading to high turbulent diffusion of momentum and over prediction of the shear layer growth. While comparing our data with the Simon et al. [11], it is noteworthy to mention that the LES modeling as well as mesh count are different in their work and that may have lead to the differences observed in the Figures 5-8. In the present work, dynamic Smagorinsky model [24-26] has been used, whereas Simon et al. [11] used a different modelling of the subgrid scale stresses, which is quite similar to constant Cs (constant smagorinsky) calculation. Secondly, we have used LDFSS (Low dissipation flux spilling scheme by Edward [27]), whereas Simon et al. [11] used modified AUSM+ and classical Roe scheme. Simon et al. [11] also found that the use of different numerical scheme may lead to different predictions of recirculation zone length.

Furthermore, it has also been observed that the core of the recirculation region (Fig. 5) is shifted to $x/R_o \sim 2$ when compared to $x/R_o \sim 1.5$ in experiments. There are two possible reasons for the same. Firstly, the experimental boundary layer thickness close to the base corner is much thicker than the computational one; that means in the computations, the amount of higher momentum fluid which enters the domain is also more. As a result, the growth of the free shear layers, which determine the length of recirculation zone ( the location where the free shear layers begin to interfere with each other near the wake axis), is larger and that's why the core in the computations might have been shifted to $x/R_o \sim 2$. The second possible reason may come from the turbulence modeling. If the eddy viscosity prediction is less, then the diffusion of free shear layer is also less, in turn effecting the growth of the free shear layer as delayed in the present case. Since, the phenomena may arise due to one of these possible reasons or due to both of them, it is not possible to identify one cause over other which requires detailed parametric investigation, may be considered as a part of future work.

**3.3 Study of Turbulence in the near wake flow field:**

In order to study the expansion of shear layer further, we have plotted the resolved Turbulent Kinetic Energy(TKE) and resolved primary Reynolds Shear Stress(RSS) profiles at three different locations in the wake region. We have also compared the same with experimental data and the data obtained by Simon et al. [11] in order to make a comparative assessment of the quality of our predictions. Following equations are used to calculate these parameters.

$$TKE = k_{average} = \frac{1}{2} \times \left( {u'_{rms}}^2 + {v'_{rms}}^2 + {w'_{rms}}^2 \right) \qquad [21]$$

$$Primary\ Reynolds\ Shear\ Stress = -\overline{u'v'}/U_1^2 \qquad [22]$$

Figure 8(a) exhibits the resolved TKE and primary RSS profiles at location right after the flow has separated where both the parameters seem to attain a very sharp peak in the thin shear layer which has been predicted by our LES study with acceptable accuracy. Further downstream, as the shear layer expands and due to higher turbulent diffusivity both TKE and primary RSS attains a more distributed profile with a higher peak value in the middle of the shear layer. At the locations further downstream (Figure 8(b-c)); LES predictions seem to under predict the shear layer thickness. Similar tendency of under prediction of the shear layer thickness has been observed in the LES study performed by Simon et al. [11] as well. From the contour plots of resolved axial turbulence intensity (Fig. 9), TKE (Fig. 10) and primary RSS (Fig. 11), it can be clearly seen that all of these parameters attain a relatively higher value in the reattachment region. From these plots it can be inferred that thickness of the shear layer gradually expands in the downstream and intensity of turbulence, turbulent kinetic energy, and magnitude of primary RSS also increases along the shear layer in the downstream direction and attains the maximum value in region where the shear layer reattaches.

**3.4 Study of Unsteady Flow Structures**

In Figure 12, the distributions of instantaneous vorticity magnitude at different times, on $z = 0$ plane have been plotted. Highly unsteady three dimensional vortical structures are observed in the wake region of the cylindrical after-body. It can be easily observed that the vortices are rapidly breaking up after passing through the reattachment region. Larger vortices are present in the downstream region where, in the recirculation region smaller vortices can be seen. From the sequence of contour plots presented in Figure 12, it can be seen how vortices are getting detached from the shear layer and forming hair-pin like structures(marked in black circle) near the recompression region and eventually breaking up as they proceed further downstream. In the recompression region, due to presence of the recompression shock system and high temperature gradient, the recompression process is not an isentropic one, thus the baroclinic contribution term $\left(\frac{\vec{\nabla}\rho\times\vec{\nabla}p}{\rho^2}\right)$in vorticity transport equation is nonzero leading to formation of convoluted vortical structures with high vorticity magnitude. This might be the reason of formation of hair-pin like structures near the recompression region.

In order to identify the 3-D coherent structures in the flow iso-surfaces of Q-criterion, is given by equation [27], has been plotted in Figure 13.

$$Q = -\frac{1}{2}\left(S_{ij}S_{ij} - \Omega_{ij}\Omega_{ij}\right) \qquad [23]$$

From Figure 13, it can be clearly seen that smaller structures lie in the recirculation region where larger structures with higher vorticity magnitude appear in the shear layer. Azimuthal coherence between the structures in the Q-criterion plot has not been observed. Yet it can be seen that there are hair-pin like 3-D vortical structures with high vorticity magnitude, which are being issued from the shear layer and seem to grow as they move in the downstream direction and eventually detach from the shear layer near the reattachment zone. Finally,further downstream behind the recompression region, these structures seem to form larger coherent structures in the far wake region. Vortical structures with similar shape are also observed in the sequence of vorticity magnitude plot and it has also been noticed that there is a sudden periodicity in the formation of these structures. Flapping of the free shear layer is also associated with periodical formation of these structures.

To find out the frequency of flapping of the shear layer, we have probed pressure and axial velocity data from a point in the shear layer and Power Spectral Density (PSD) analysis of that data has revealed sharp peaks at frequency of approximately 2.5 kHz(Figure 14). Figure 15 depicts the variation of pressure over one cycle of 360$^o$ at that point. Further to analyze the evolution of flow structure over one cycle of the observed frequency, streamlines at 10 different time-steps representing 10 phases with phase difference of 36$^o$ have been plotted (Figure 16). Each of these phases has been averaged over 20 cycles to eliminate the high frequency oscillation of the flow field.

In these figures (Fig. 16), it can be observed that the vortices generated upstream in the shear layer tend to gradually grow and move towards downstream along the shear layer. In pahse-1, a newly generated vortex in the lower shear layer closer to the base surface can be observed which seems to grow in phase-2 and eventually merge with the bigger vortex in phase-3. The larger vortex core in the shear layer moves toward the reattachment zone further in phase-4 and phase-5;by that time another vortex starts forming near the base surface (phase-7) and the similar set of events are repeated again. Similar kind of dynamic behavior can be observed in the upper shear layer as well. From these events it can be inferred that the shear layer instabilities form vortices upstream in the shear layer which gradually grows and travels downstream along the shear layer eventually break-up at the recompression region leading to formation of the hair-pin like vertical structures seen earlier in the sequence of contour plot of vorticity magnitude on z=0 plane. To further study these flow structures, we have performed proper orthogonal decomposition on a 2-D central plane (z=0).

### 3.5 POD analysis

For POD analysis data is extracted from a rectangular region in Z=0 plane along the centerline. After that the data is extrapolated on 2D uniform grid, three different grid resolutions($100 \times 100, 200 \times 200$ $and$ $400 \times 400$) have been considered to study the effect of grid resolution on the POD results. Three different time-steps (time difference between two consecutive snap-shot) respectively $6 \times 10^{-5}$ $seconds, 8 \times 10^{-5}$ $seconds$ $and$ $10 \times 10^{-5}$ $seconds$ are also considered to study the effects of varying time-step on the POD results.

From Figure 17, it is evident that for both of energy based and enstropy based POD, the Eigen values obtained from the mentioned grid resolutions seem to have a little difference and they seem to converge at the highest grid resolution. Again Figure 17 suggests that the first and second choice for the time-step seem to cause a little discrepancy in the Eigen values whereas the second and the third choice have rendered Eigen values which are in better agreement with each other. For this current study the grid resolution of $400 \times 400$ and time-step of $10 \times 10^{-5}$ $seconds$ have been selected.

Since enstropy is more related to the first order derivative of fluctuating velocity field, the basis functions forming the enstropy field are different from the basis functions forming the velocity field. Thus it is expected that there are going to be some differences in the POD modes obtained from energy based POD and enstropy based POD. If the normalized Eigen values obtained from energy based POD and enstropy based POD are compared, it can be seen that the distribution of normalized Eigen values are little different. Figure 18 (a-j) exhibits the first ten modes obtained from energy based POD where the contour indicated distribution of fluctuating component of energy (obtained from first three components of the POD mode which correspond to the fluctuating velocity components) across the domain and the vectors are formed from the in-plane components of fluctuating velocity field (length of the vectors shown are not in proportion with the magnitude of the in-plane components of fluctuating velocity field). Figure 18 (right column) also depicts the distribution of first tenenstropy modes over the domain. Though the contour plots of first five POD modes obtained from energy and enstropy based POD do not exhibit exact similar mode distributions across the plane; however, it is noteworthy to mention that there are some intrinsic similarities.Both of the energy and enstropy based POD analysis have revealed that the shear layer and recompression zone play a major role in the fluid dynamic processes occurring in the wake region.From the POD shape distributions it is evident that the highest concentration of energy and enstropy occur in these regions.This observation is also in accord with the previous conjectures drawn from contour plot of resolved turbulent kinetic energy, primary Reynolds shear stress and axial turbulence intensity. None of the energy and enstropy based POD modes exhibit presence of bi-dimensional coherent structures issued due to Kelvin-Helmholtz instability in the shear layer; but the presence of some transient flow structures in the shear layer has been captured through the POD modes. The$1^{st}$, $4^{th}$ and $5^{th}$ (Figure 18) POD mode obtained from energy based POD revealedvortical structures in the shear layer which has been seen earlier in the phase averaged streamline plots (Figures 16), whereas the $2^{nd}$ (Figure 16) mode exhibits mushroom like structures which are being issued form the recompression region as seen earlier in the vorticity contour plot of unsteady. In $2^{nd}$ and $3^{rd}$ modes obtained from energy based POD, presence of a large vorticalstructure can be seen near the base which assumes opposite direction of rotation in these modes. But from the contour plots it is evident that these structures contain less energy that the structures observed in the shear layer. Presence of the reattachment shock system can also be observed in the energy based POD modes. In 6$^{th}$, 7$^{th}$ and 8$^{th}$ mode, presence of larger vortical in the far wake region,behind recompression zone can be observed. Where the 7$^{th}$ mode depicts the presence of these structures towards the lower shear-layer, the 8$^{th}$ mode showsthem to be positioned towards the upper shear layer. It is also noticeable that 7$^{th}$ and 8$^{th}$ modes are having relatively less difference in the Eigen values, suggesting they are having almost equal contribution towards the flow-field.POD modes clearly exhibit presence of large vortical structuresin the reattachment region, while structures seen inside the recirculation region seem to be present and also contain less energy. Though periodic shedding of vortices have not been observed, presence of these vortical structures in the reattachment zone and the presence of smallervortical structures in the shear layer are in accord with the earlier observations we made in the vorticity magnitude contour plots (Figure 12) and the phase averaged streamline plots (Figure 16).

**4. CONCLUSIONS**

A high Reynolds number supersonic base flow is numerically investigated using RANS and LES. It is found that the LES is able to predict the features of mean flow field and the turbulent properties of the flow field successfully and the predictions are found to be satisfactory. Unsteady flow structures such as large eddies in the downstream region, smaller vortices inside the recirculation region, mushroom like structures near the reattachment region are also properly captured by the LES. On the other hand the $k$-$\varepsilon$ model fails to predict the flow field in the near wake region. Presence of strong shear layer and vortices interacting with the recompression shock system induces strong unsteadiness in the flow. This might be the reason why standard $k$-$\varepsilon$ method fails to predict the turbulent mixing properly across the shear layer and eventually renders wrong prediction of the recirculation region. From the detailed analysis of unsteady, data it has been found out that in the shear layer instabilities in form of vortical structure are generated which travel further downstream along the shear layer and eventually detach from the shear layer near the recompression region forming the mushroom or hairpin like structures. Though the POD analysis of the data obtained from a 2D plane along the centerline has not shown the presence of coherent structures, but the POD modes have confirmed the presence of vortical structures in the shear layer and the mushroom like structures in the reattachment region.

**ACKNOWLEDGMENTS:**

The authors would like to acknowledge High Performance Computing (HPC) Facility at IIT Kanpur (www.iitk.ac.in/cc). Also, we would like to thank Dr. A.C.Mandal of AE, IITK for his valuable inputs regarding the data analysis.

**REFERENCES:**

[1] J.L. Herrin, J.C. Dutton, Supersonic Base Flow Experiments in the Near Wake of a Cylindrical Afterbody, AIAA J. 32(1) (1994) 77–83.

[2] J. Sahu, Numerical Computation of Supersonic Base Flow with Special Emphasis on Turbulence Modeling, AIAA J. 32(7) (1994) 1547–1549.

[3] C.C. Chuang, C.C. Chieng, Supersonic Base-Flow Computation Using Higher-Order Closure Turbulence Models, J. Spacecraft Rockets 33(3) (1996) 374–380.

[4] R. Benay, P. P. Servel, Two-Equation $k$−$\sigma$ Turbulence Model: Application to a Supersonic Base Flow, AIAA J. 39(3) (2001) 407–416.

[5] J.L. Papp, K.N. Ghia, Application of the RNG Turbulence Model to the Simulation of Axisymmetric Supersonic Separated Base Flow, 2001, AIAA Paper 2001-0727.

[6] D. De Feo, S. Shaw, S., Turbulence Modeling and Supersonic Base Flows, 45th AIAA Aerospace Sciences Meeting and Exhibit, 2007.

[7] M. Dharavath, P.K. Sinha, D. Chakraborty, Simulation of supersonic base flow: effect of computational grid and turbulence model, Proc. Inst. Mech. Engineers, Part G: J. Aerospace Eng. 224 (2010) 311-319.

[8] J.R. Forsythe, K.A. Hoffmann, R.M. Cummings, K.D. Squires, Detached-Eddy Simulation with Compressibility Corrections Applied to a Supersonic Axisymmetric Base Flow, J. Fluids Eng. 124(4) (2002) 911–923.

[9] S. Kawai, K. Fujii, Computational Study of a Supersonic Base Flow Using LES/RANS Hybrid Methodology, 2004 AIAA Paper 0068.

[10] S. Kawai, K. Fujii, Computational Study of Supersonic Base Flow Using Hybrid Turbulence Methodology, AIAA J. 43(6) (2005) 1265-1275.

[11] F. Simon, S. Deck, P. Guillen, P. Sagaut, Reynolds Averaged Navier–Stokes/Large-Eddy Simulations of Supersonic Base Flow, AIAA J. 44(11) (2006) 2578-2590.

[12] G. Rodebaugh, K. Brinckman, S. Dash, DDES of Aeropropulsive Flows Based on an Extended k-ϵ RANS Model, 51st AIAA Aerospace Sciences Meeting including the New Horizons Forum and Aerospace Exposition, 2013.

[13] J.R. Forsythe, D.H. Hine, P.H. Benjamin, J.P. Laiosa, T.C. Shafer, Fundamental Physics Validation Using HPCMP CREATETM-AV Kestrel: Part II, 52nd Aerospace Sciences Meeting, 13-17 January 2014, Maryland.

[14] D. Luo, C. Yan, X. Wang, Computational study of supersonic turbulent-separated flows using partially averaged Navier-stokes method, Acta Astronautica 107 (2015) 234-246.

[15] R.D. Sandberg, H.F. Fasel, Direct Numerical Simulations of Transitional Supersonic Base Flows, 2005 AIAA Paper 2005-98.

[16] K.Y. Chien, Prediction of Channel and Boundary-layer Flows with Low-Reynolds Number Turbulence Model, AIAA J. 20(1) (1982) 33-38.

[17] A. De, S. Acharya, Large Eddy Simulation of Premixed Combustion with a Thickened-Flame Approach, ASME J. Eng. Gas Turbines Power 131(6) (2009) 061501.

[18] A. De, S. Acharya, Large Eddy Simulation of a Premixed Bunsen flame using a modified Thickened-Flame model at two Reynolds number, Combust. Sci. Technol. 181(10) (2009) 1231-1272.

[19] A. De, S. Zhu, S. Acharya, An experimental and computational studyof a swirled stabilized premixed flame, ASME J. Eng. Gas Turbines Power 132 (7) (2010) 071503.

[20] A. De, S. Acharya, Parametric study of upstream flame propagation in hydrogen-enriched premixed combustion: Effects of swirl, geometry and premixedness. Int. J. Hydrogen Energy 37 (2012) 14649-14668.

[21] A. De, S. Acharya, Dynamics of upstream flame propagation in a hydrogen enriched premixed flame, Int. J. Hydrogen Energy 37(2012) 17294-17309.

[22] N. Arya, R.K. Soni, A. De, Identification of coherent structures in a supersonic flow past backward facing step, AIP Conf. Proc. 1648 (2015) 030037.

[23] J.H. Ferziger, M. Peirc, Computational methods for Fluid Dynamics, Springer, 2012

[24] M. Germano, U. Piomelli, P. Moin, W.H. Cabot, A dynamic subgrid-scale eddy viscosity model, Phys. Fluids 3 (1991) 1760.

[25] D.K. Lilly, A proposed modification of the Germanos subgrid scale closure method, Phys. Fluids 4(3) (1992) 633–635.

[26] J. Smagorinsky, General circulation experiments with the primitive equations. I: The basic experiment, Month Weather Rev. 91(1963) 99-104.

[27] J.R. Edwards, A low-diffusion flux-splitting scheme for Navier Stokes calculations, Comput. Fluids 26(1997) 635-659.

[28] J.M. Weiss, W.A. Smith, Preconditioning applied to variable and constant density flows, AIAA J. 33(1995) 2050-2057.

[29] J.L. Lumley, The structure of inhomogeneous turbulence, Proc. Int. Coll. Radio Wave Propagation ( ed. A. M. Yaglom, V. I. Tatarski), 1997 166-178.

[30] L. Sirovich, Turbulence and the dynamics of coherent structures. Part 1: coherent structures, Quart Appl. Math. 45 (1987) 561–571.

[31] J.L. Lumley, A. Poje, A Low-dimensional models for flows with density fluctuations, Phys. Fluids 9(7) (1997) 2023-2031.

[32] A. Qamar, S. Sanghi, Steady supersonic flow-field predictions using proper orthogonal decomposition technique, Comput. Fluids 38(2009) 1218–1231.

[33] I. Celik, Z. Cehreli, I. Yavuz, Index of resolution quality for large eddy simulation, J. Fluid Eng. 127 (2005) 949-958.

[34] K. Akselvoll, P. Moin, Large-eddy simulation of turbulent confined coannular jets, J. Fluid Mech. 315 (1996) 387-411.

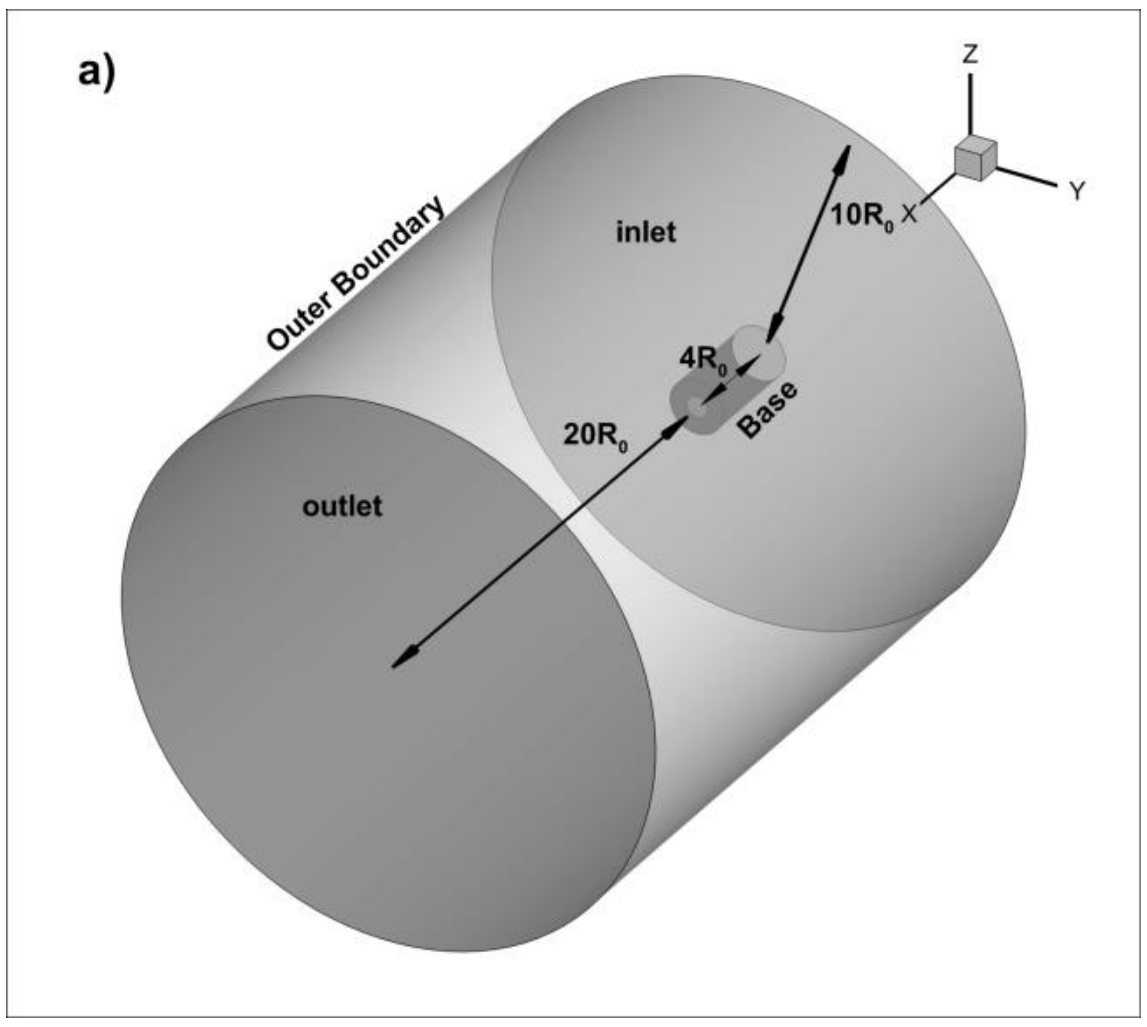

Figure 1(a): Computation Domain

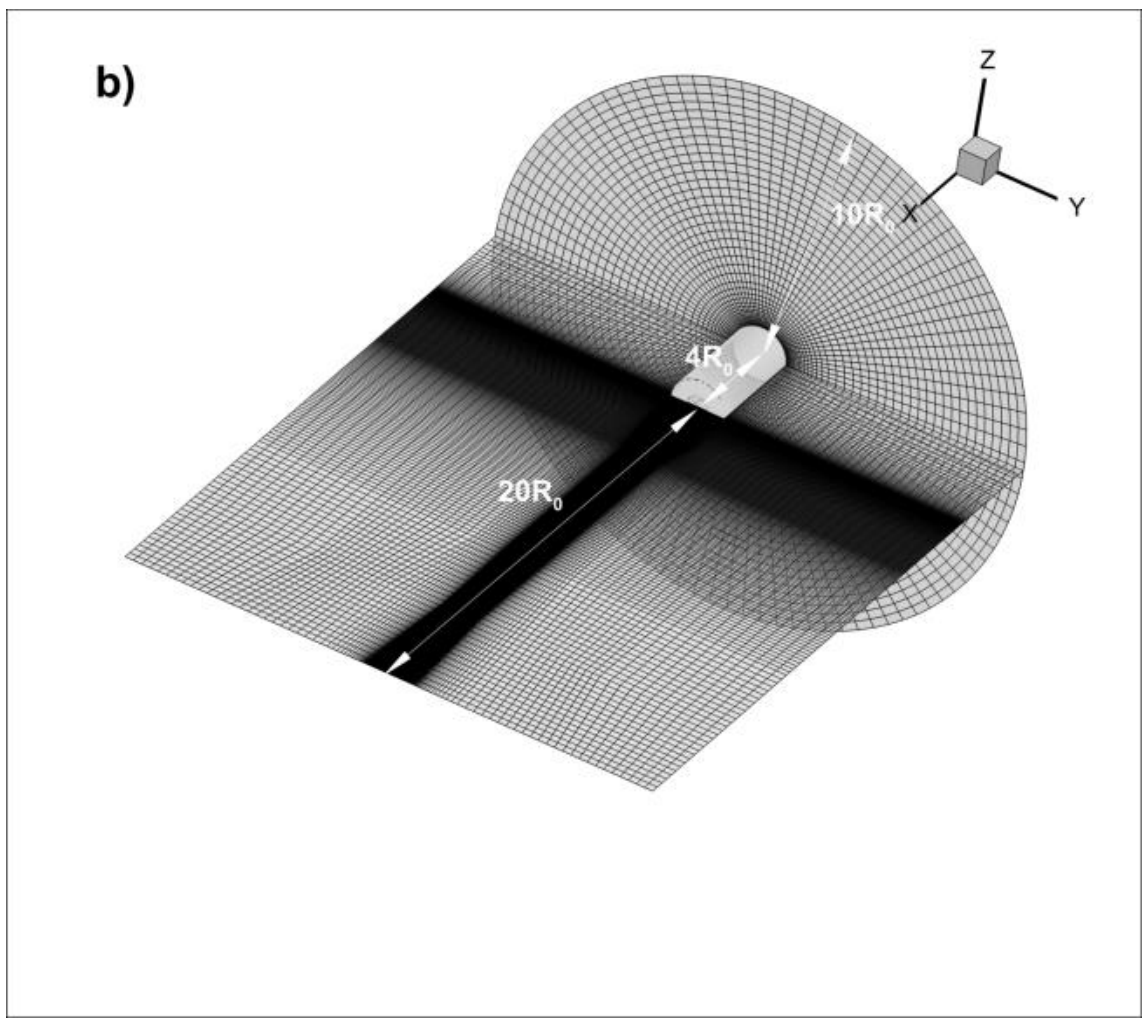

Figure 1(b): Meshing Strategy

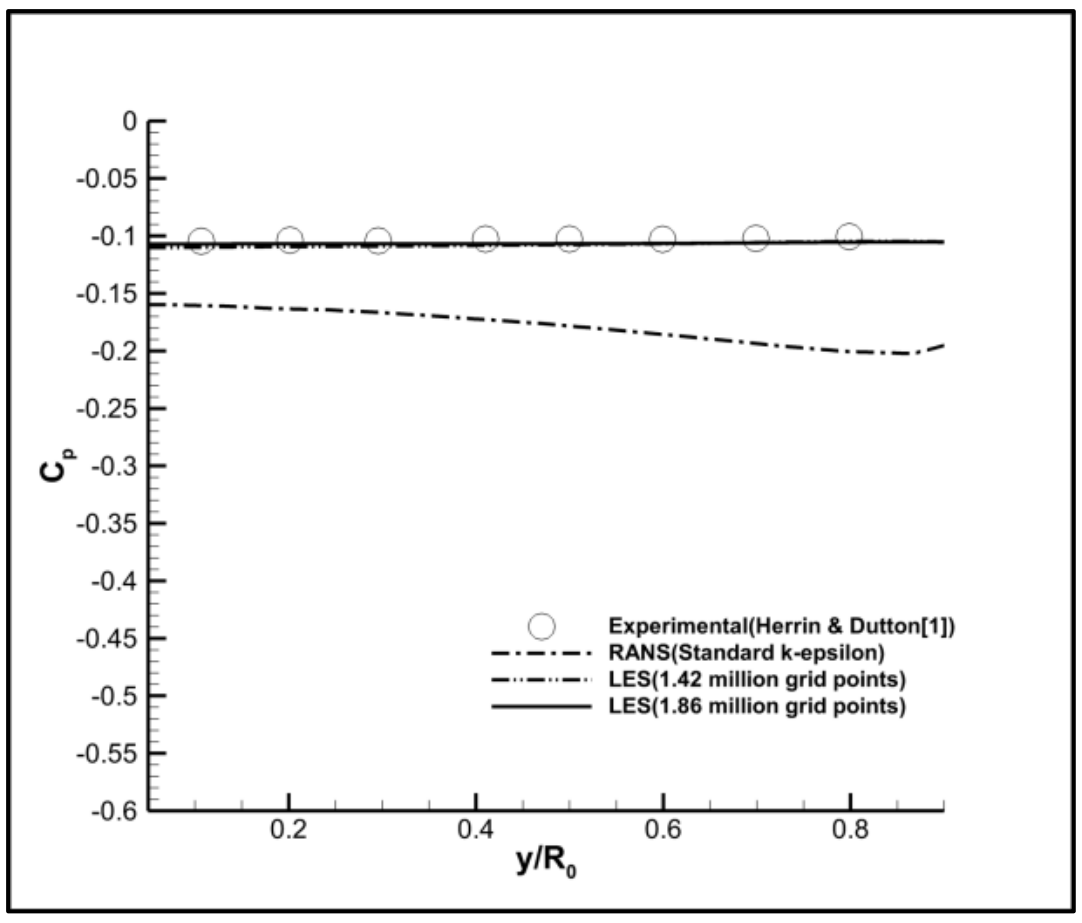

Figure 2(a): Distribution of Base pressure co-efficient on base surface

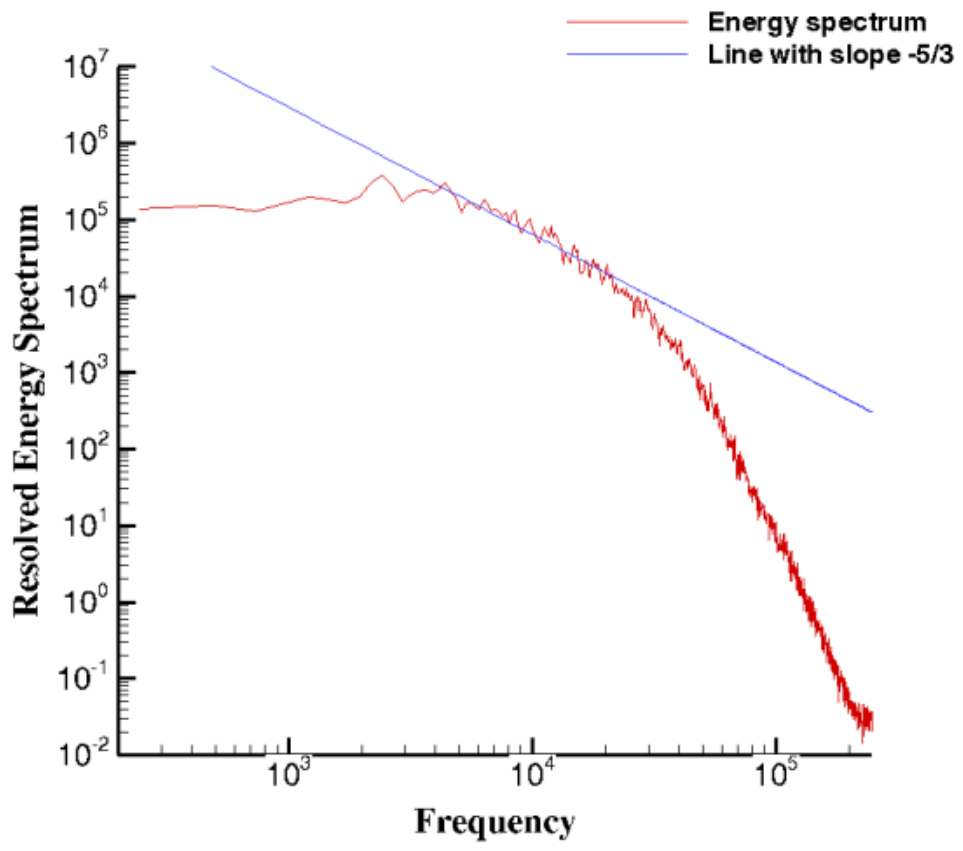

Figure 2(b): Resolved Energy Spectrum

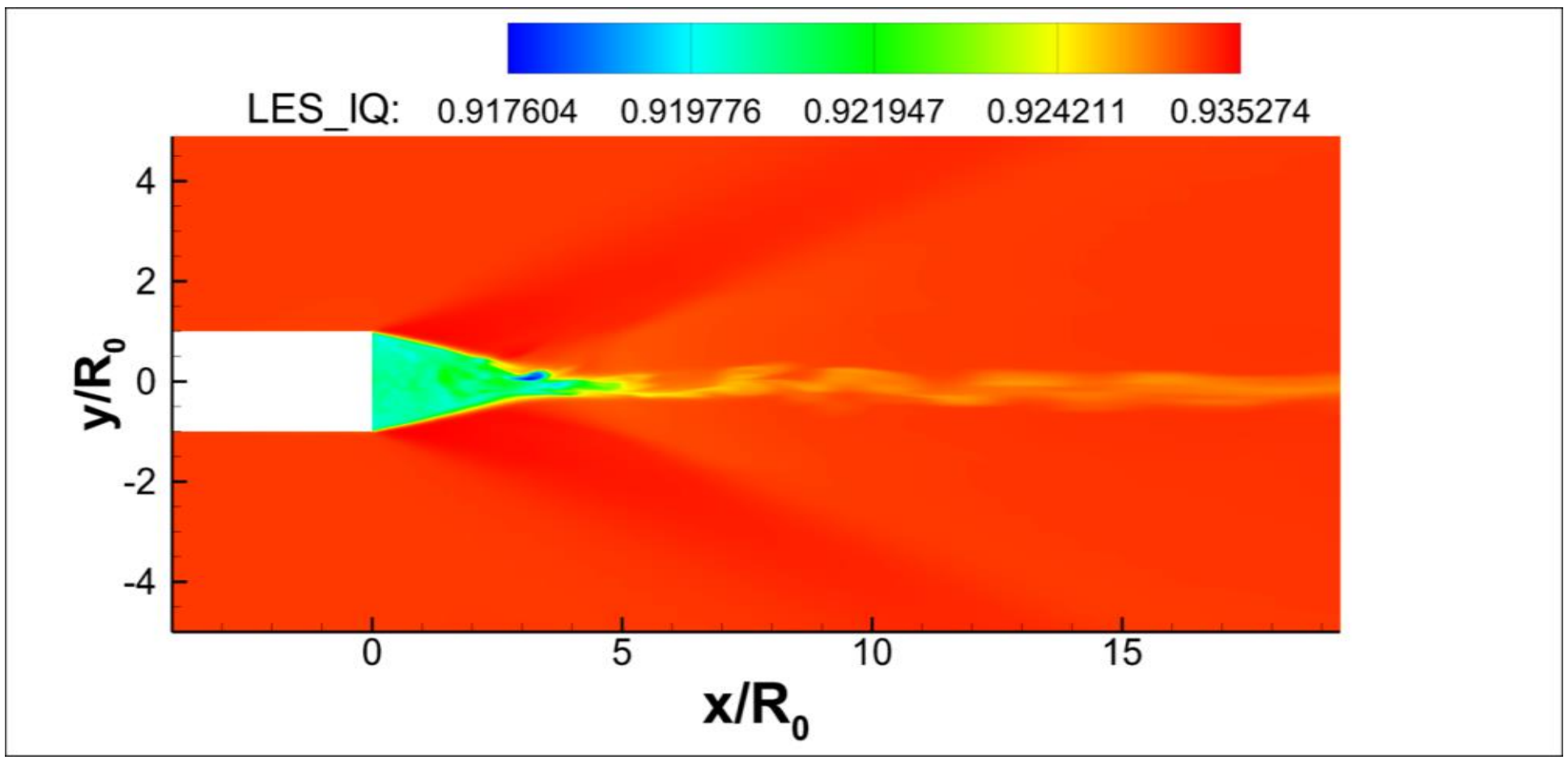

Figure 3: Resolution of Grid using LES quality criteria [33]

a)

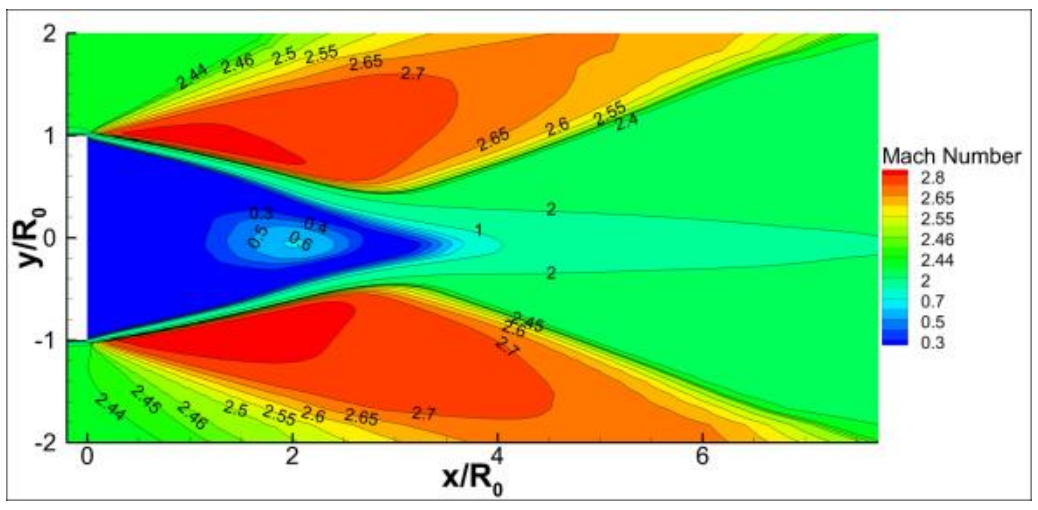

Figure 4(a): Contour Plot of time averaged Mach Number obtained from Large Eddy Simulation

b)

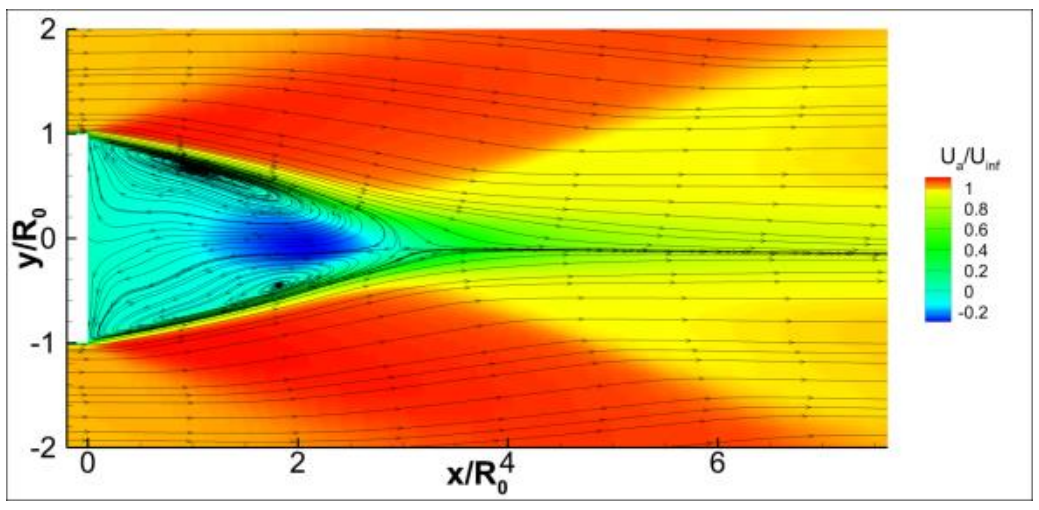

Figure 4(b): Contour Plot of time averaged axial velocity along with stream lines obtained from Large Eddy Simulation

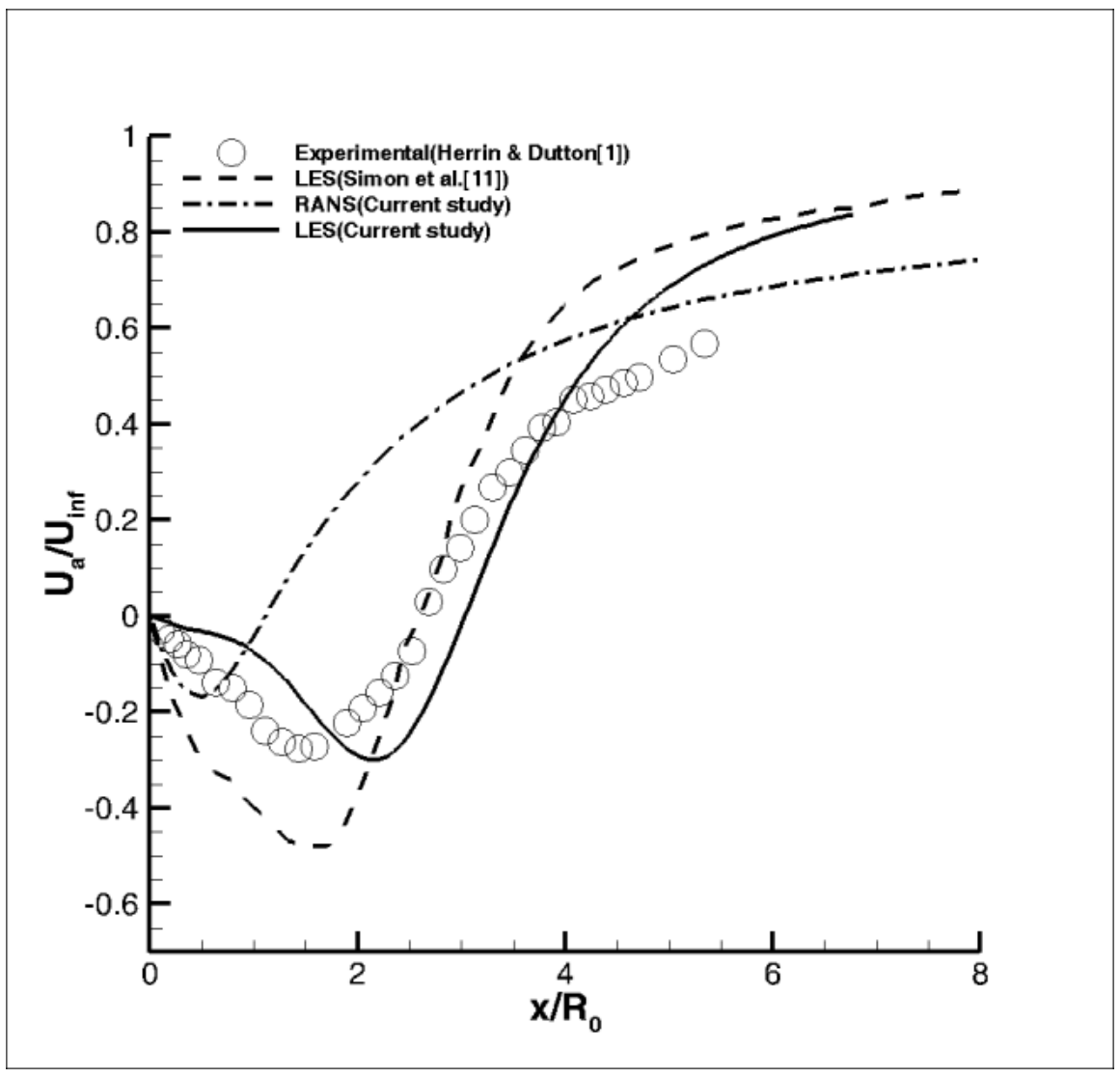

Figure 5: Time averaged axial velocity distribution along the axis of the wake

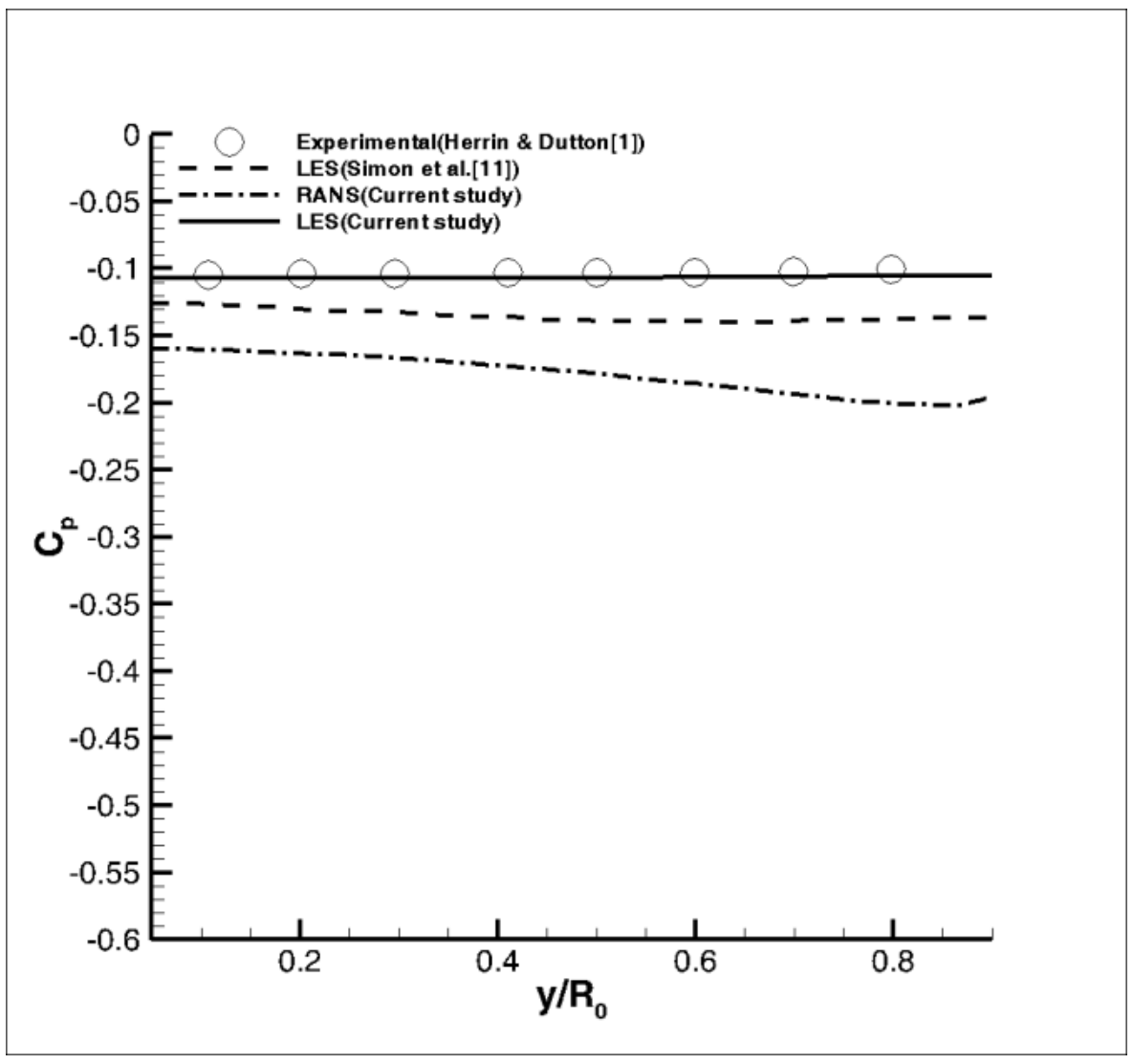

Figure 6: Time averaged pressure distribution along radial direction on the base surface

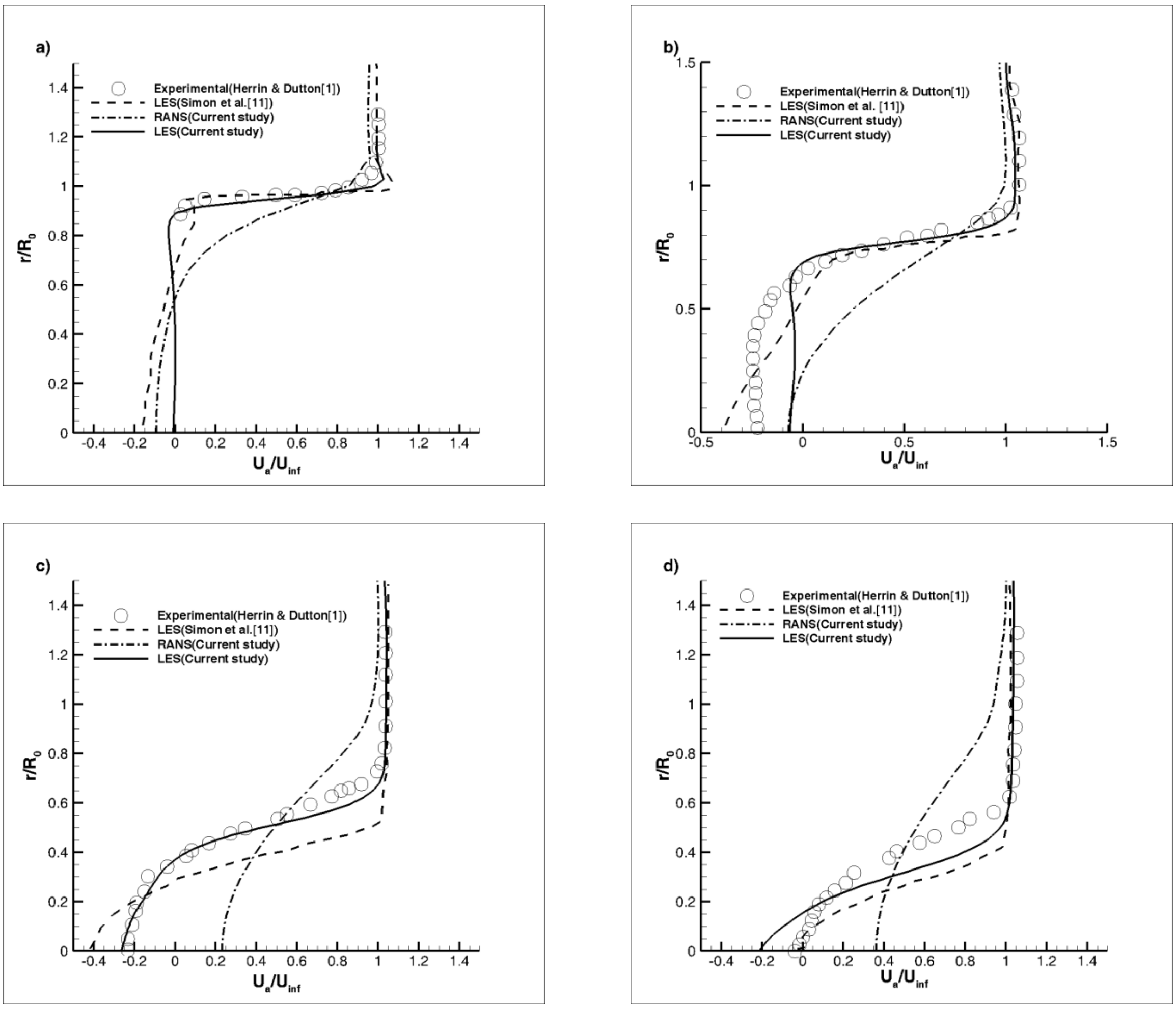

Figure 7: Time averaged axial velocity (mean) profile along radial direction at different axial locations (downstream of the base): (a) $X/R_0$=0.1575 (b) $X/R_0$=0.9449 (c) $X/R_0$=1.8898 (d) $X/R_0$=2.5197

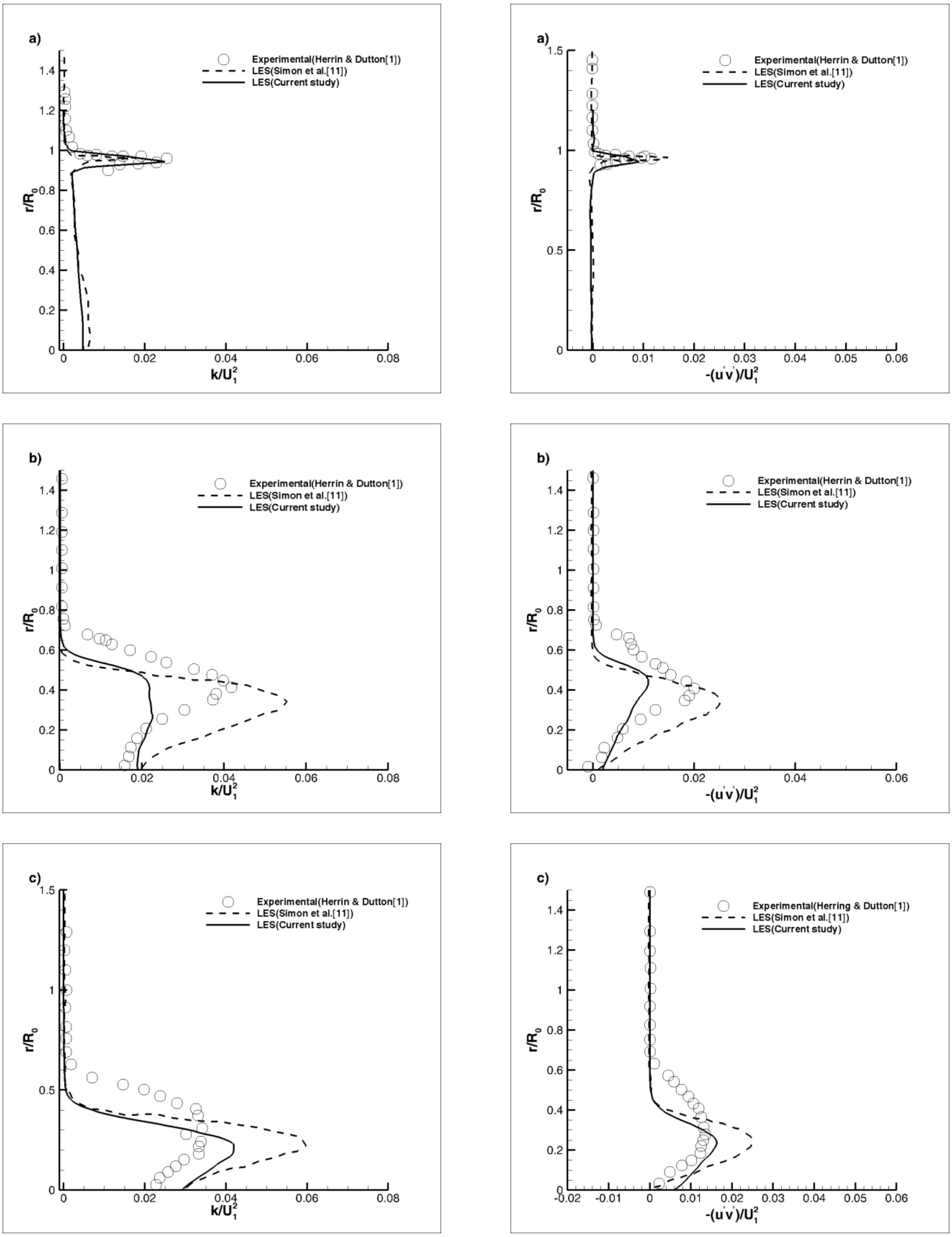

Figure 8: Time averaged radial profiles of TKE (left) and RSS (right) at different axial locations: (a) $X/R_0=0.1575$ (b) $X/R_0=1.8898$ (c) $X/R_0=2.5197$

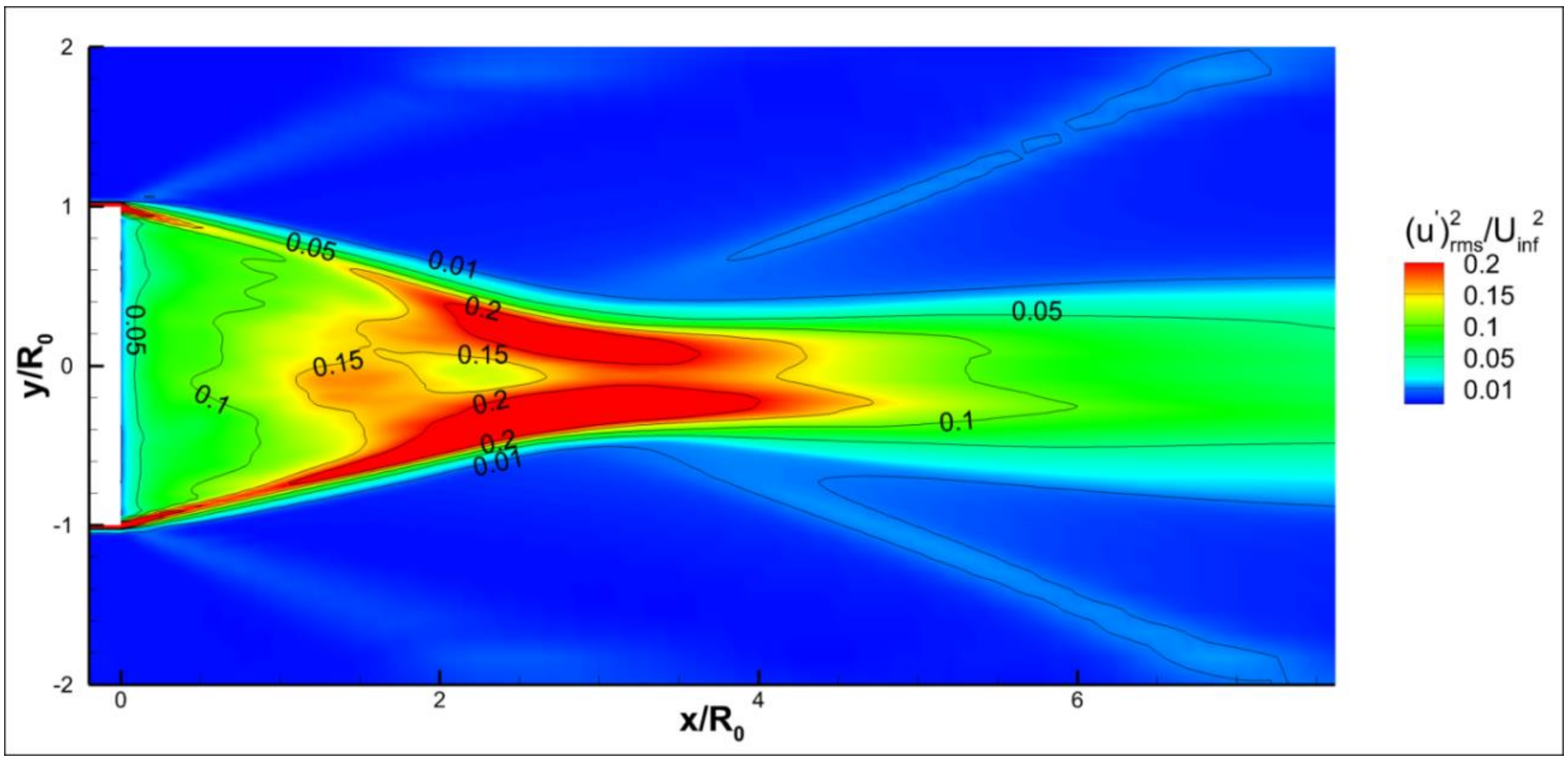

Figure 9: Contour plot of axial turbulence intensity at z=0 plane

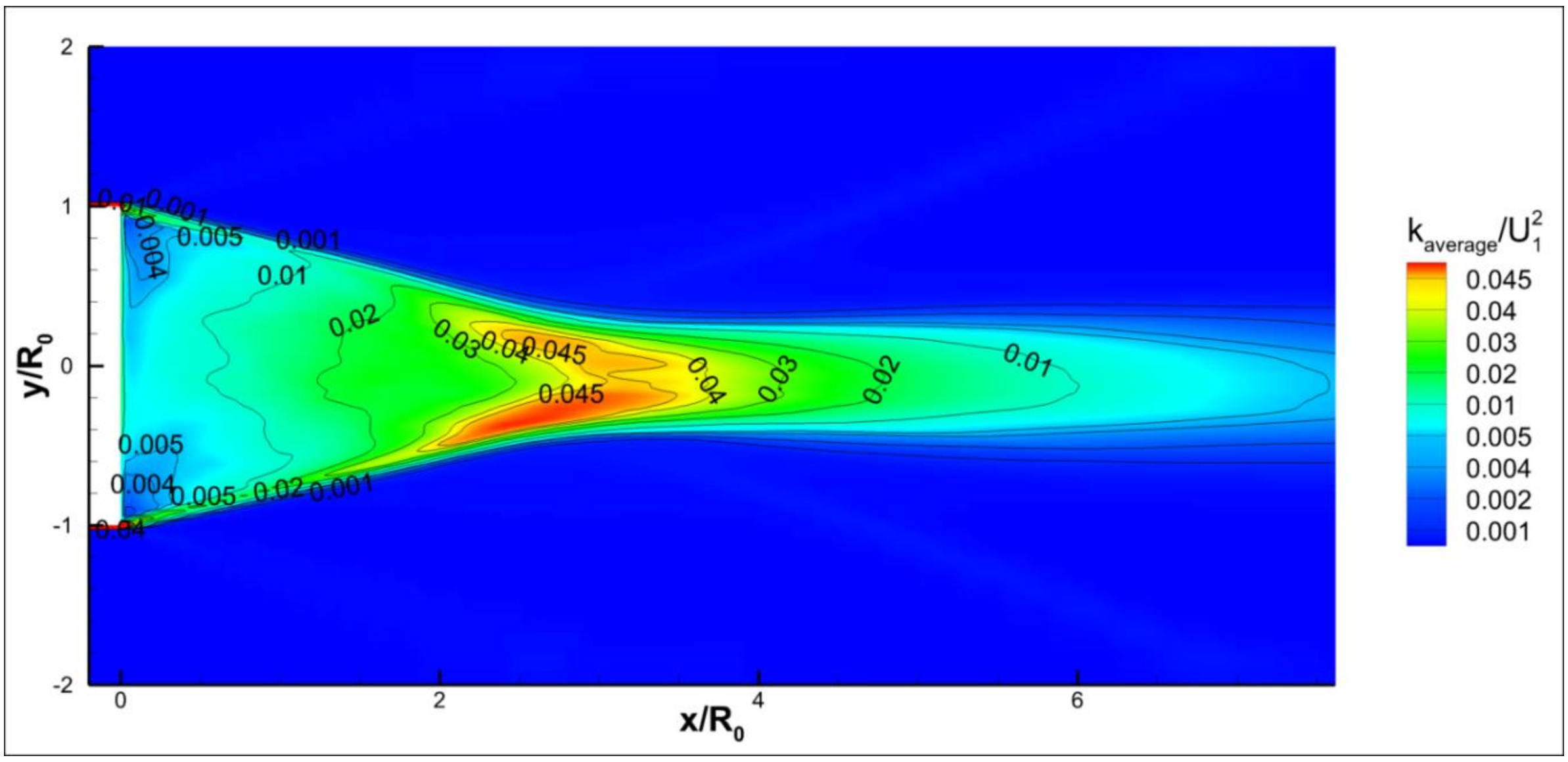

Figure 10: Contour plot of resolved TKE at z=0 plane

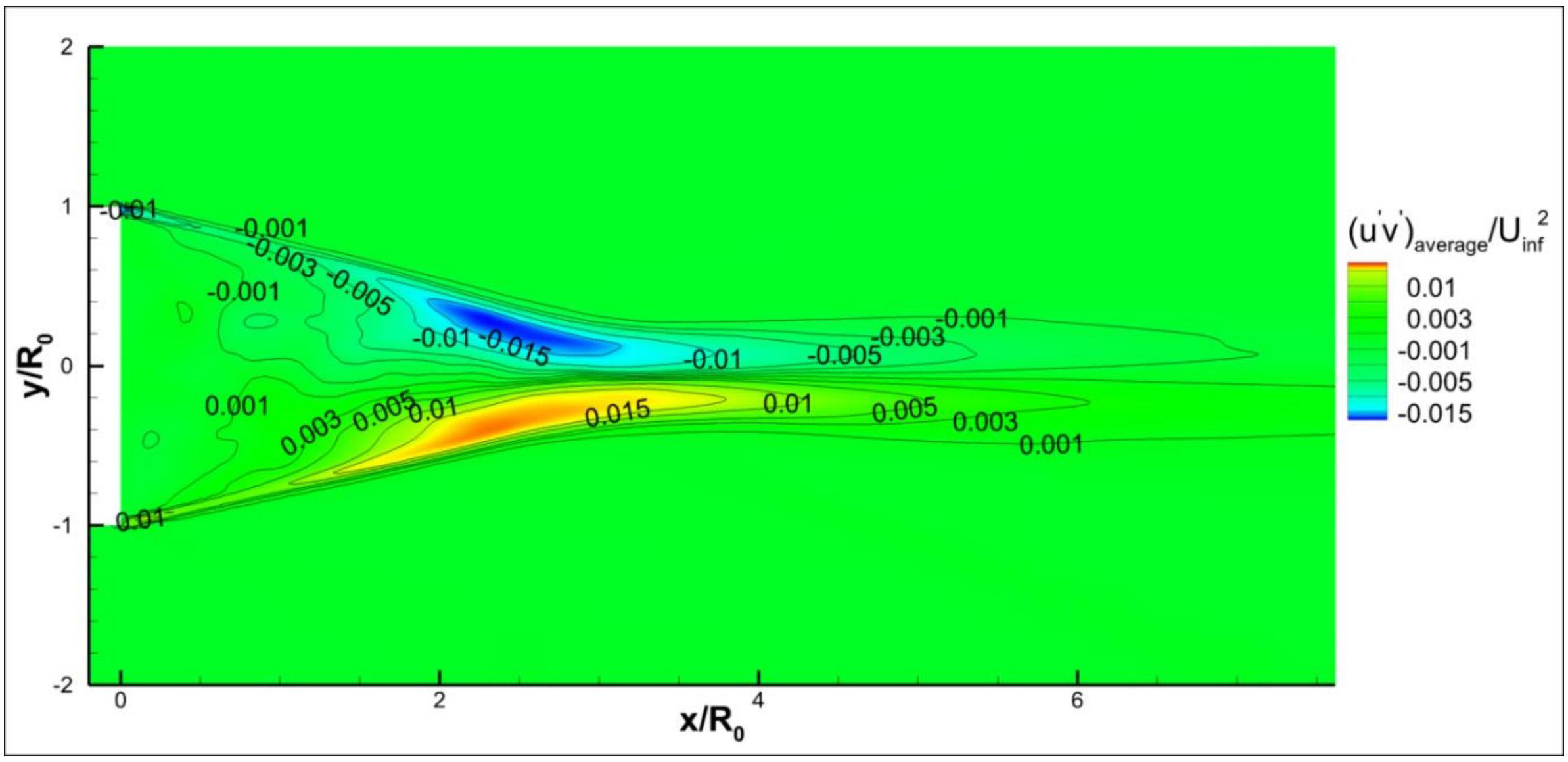

Figure 11: Contour plot of resolved primary RSS at z=0 plane

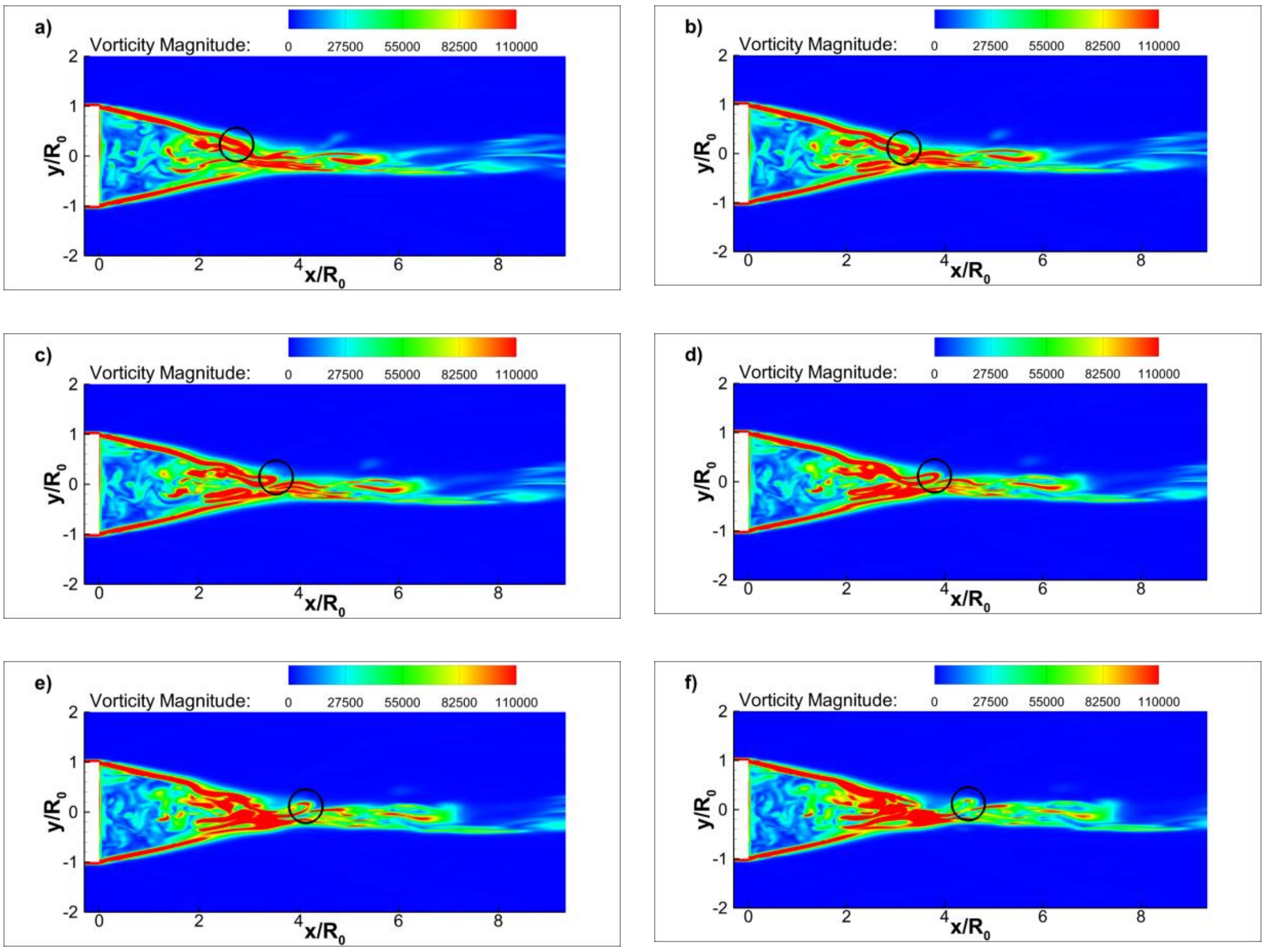

Figure 12: Vorticity contour at different time instants: (a) t=0.006326s (b)t=0.006328s (c) t=0.006330s (d) t=0.006332s (e) t=0.006334s (f) t=0.006336s

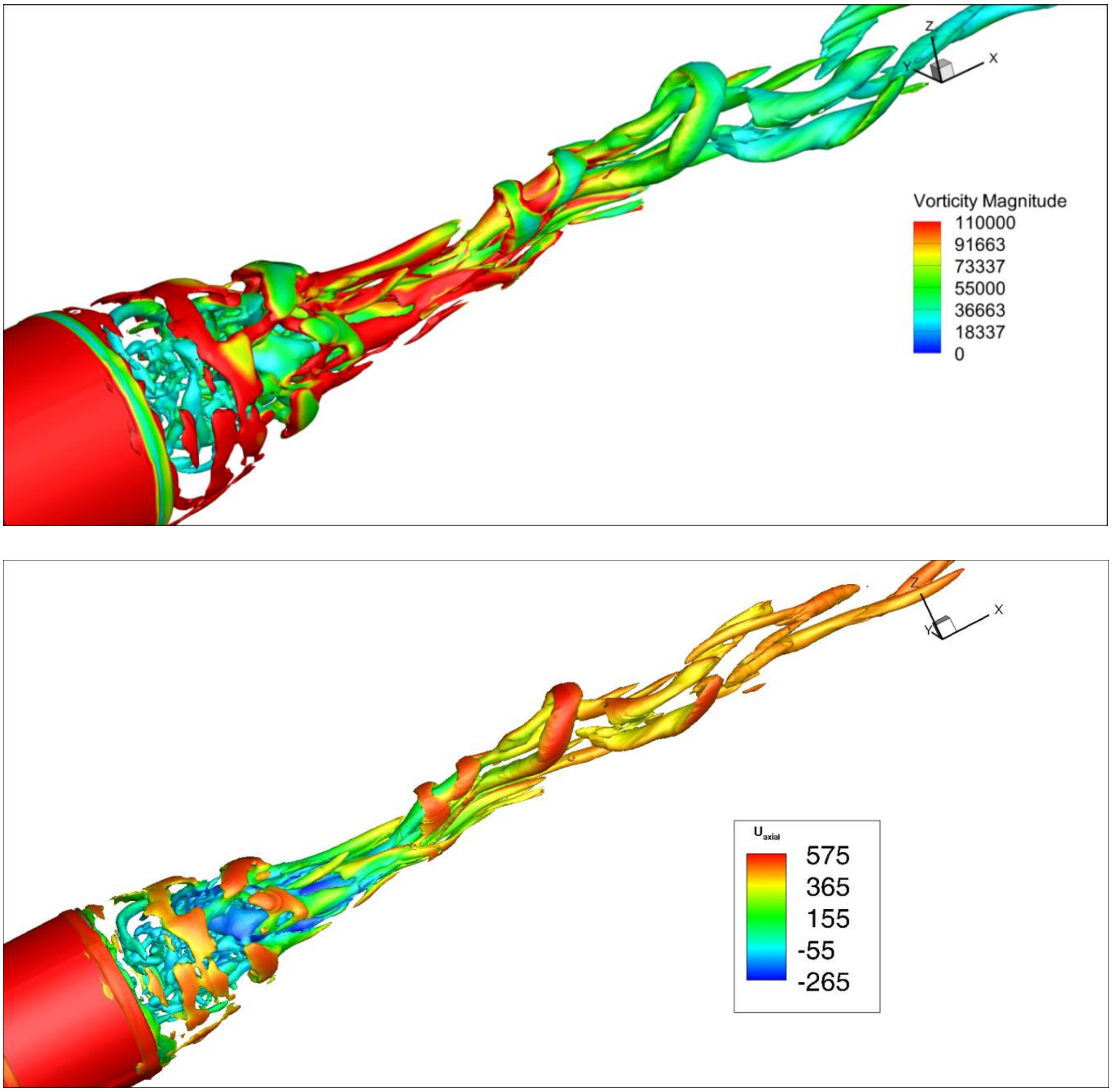

Figure 13: Iso-surface of Q criterion colored by: (a) vorticity magnitude (top) and (b) axial velocity (bottom)

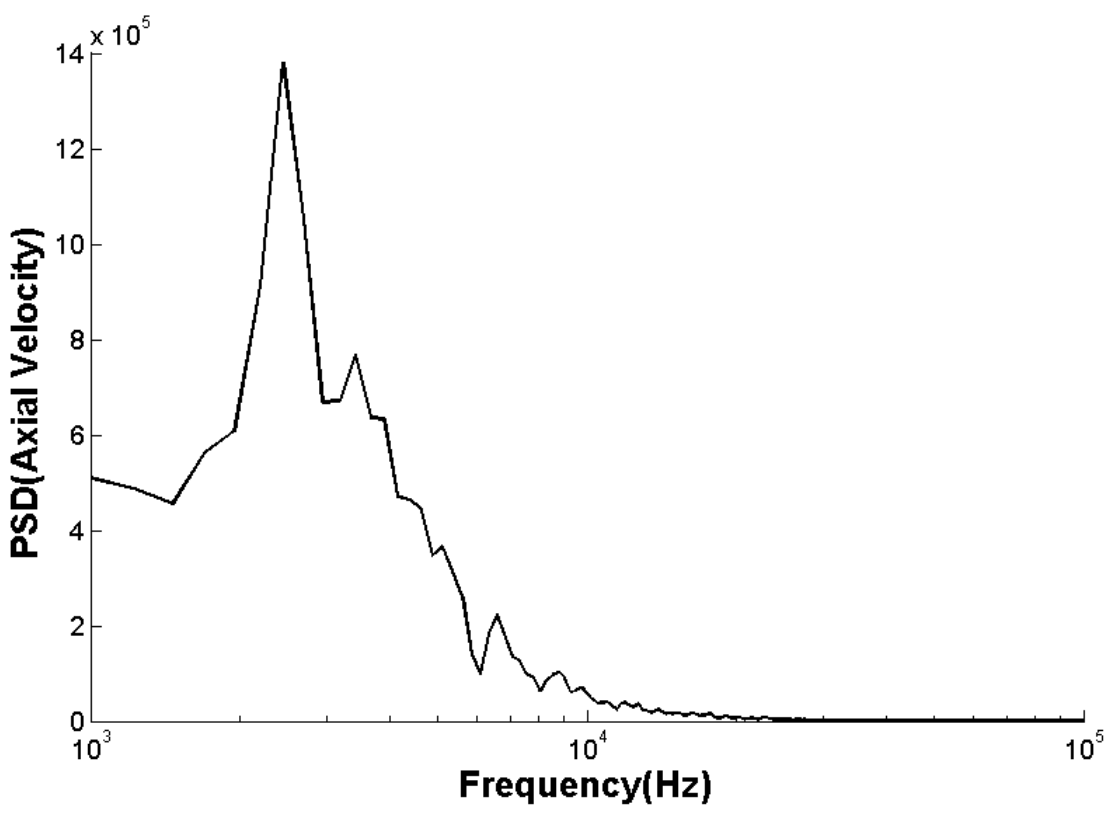

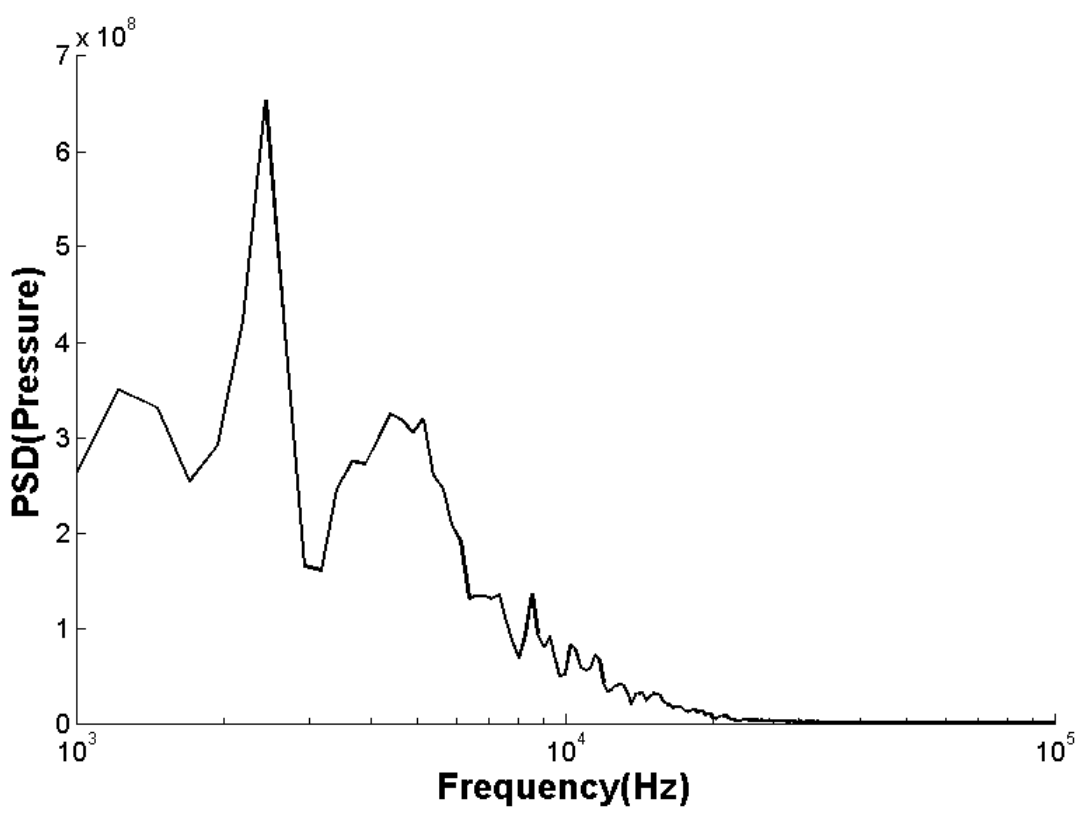

Figure 14 (a): PSD of axial velocity at a point in shear layer

Figure14(b): PSD of pressure at a point in shear layer

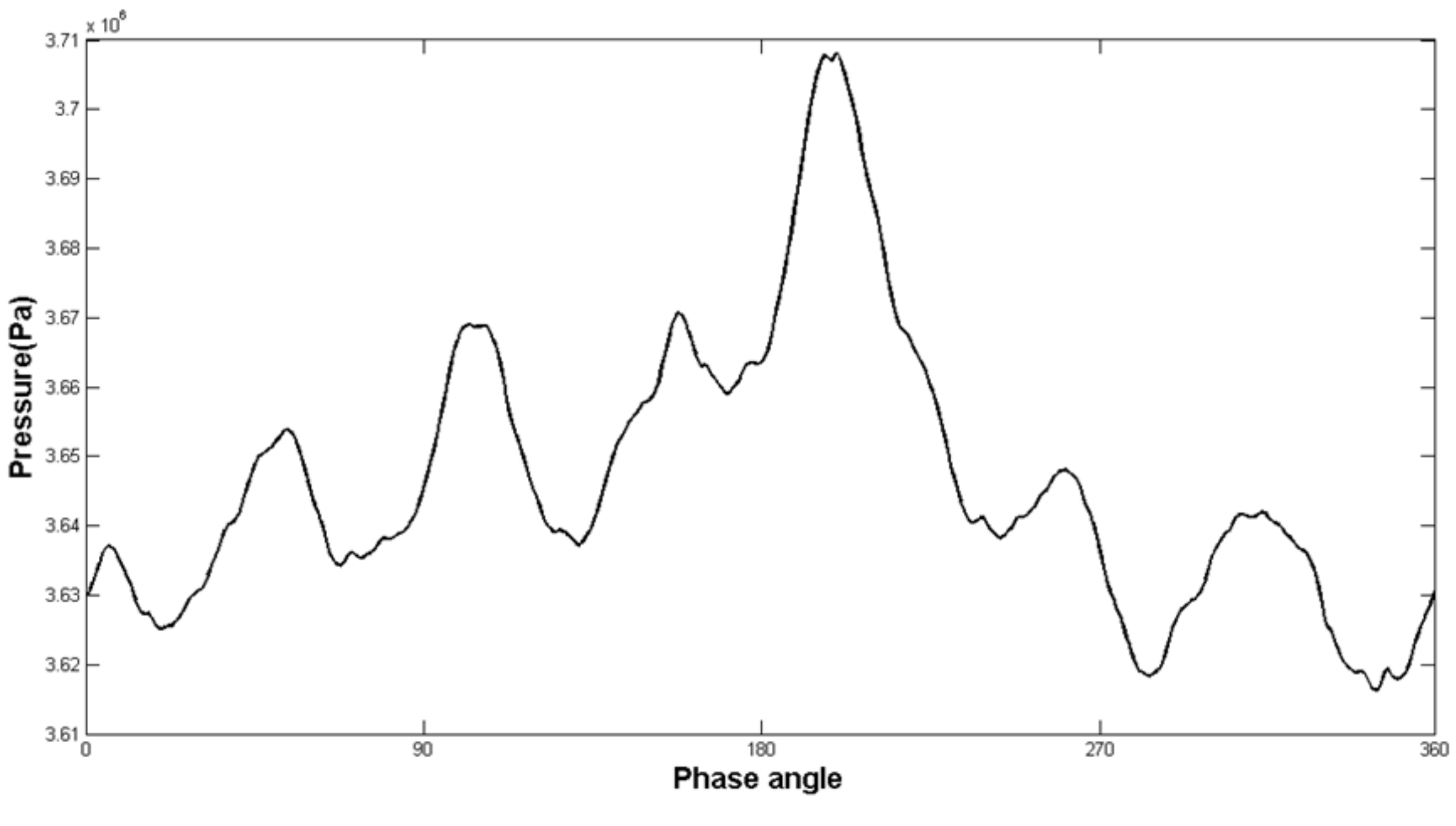

Figure 15: Pressure variation across the period of 1 complete of cycle of the observed frequency

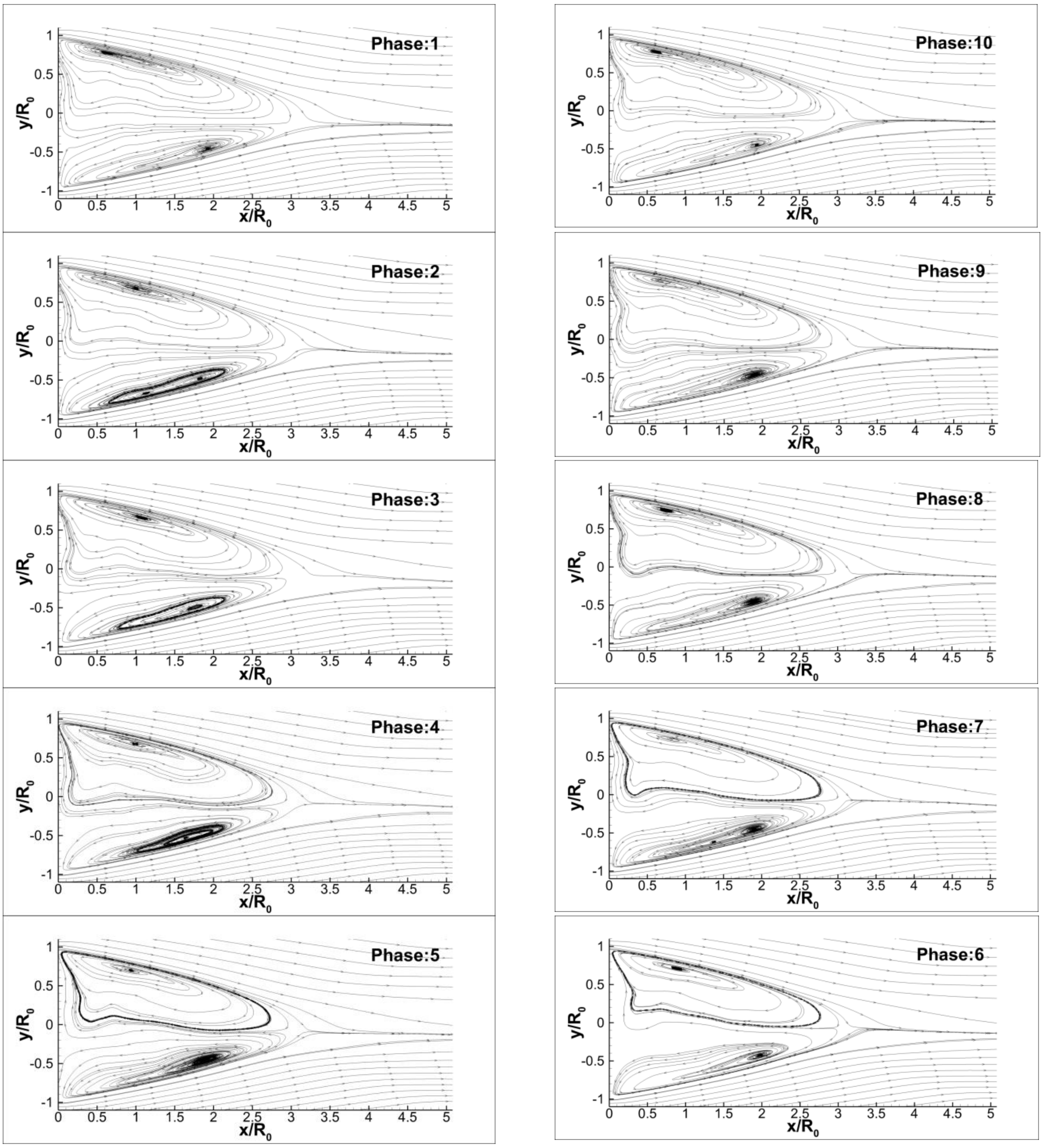

Figure 16: Phase averaged plots over a complete cycle of the observed frequency: phase1-phase10 (anticlockwise starting from left corner to right corner)

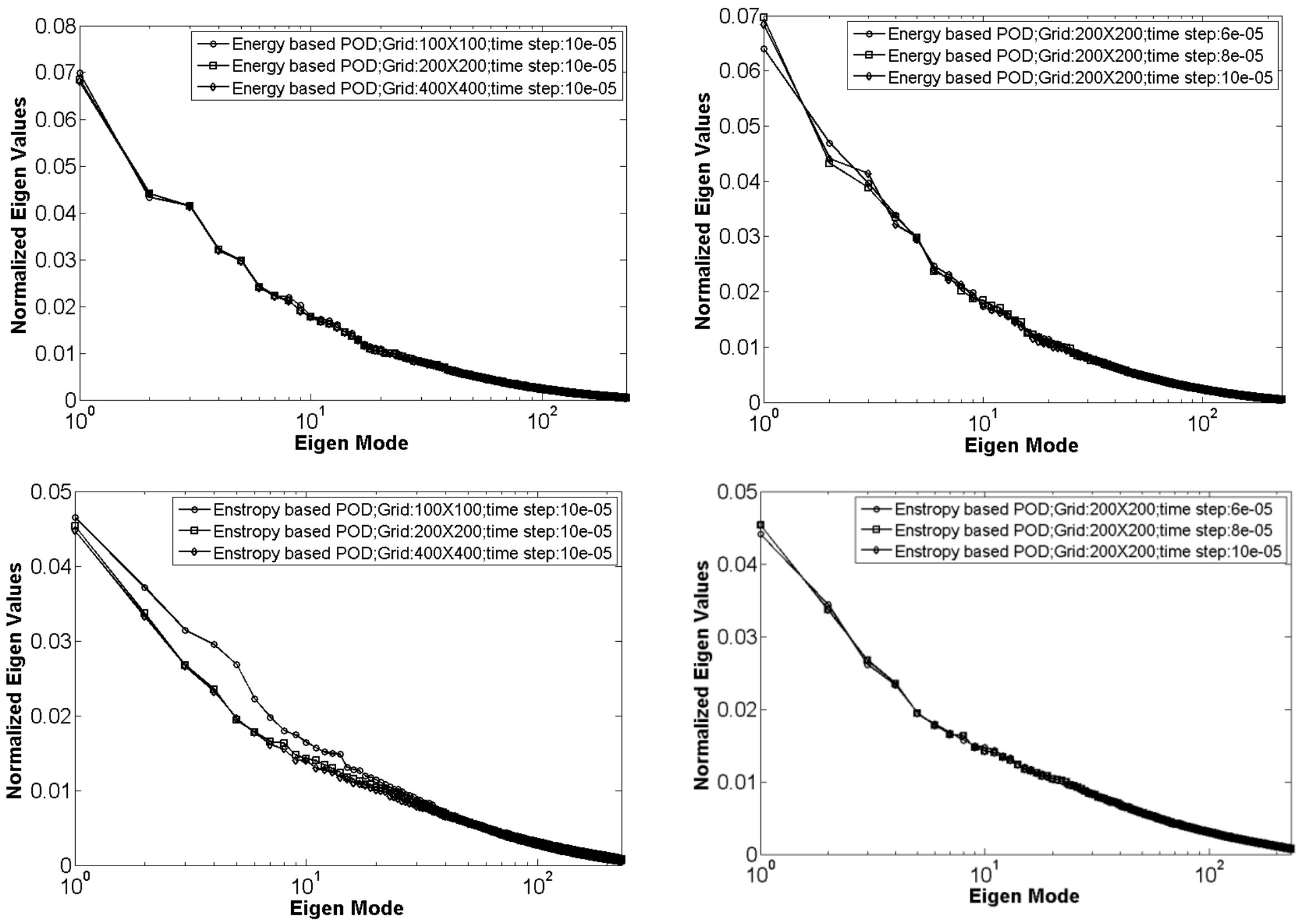

Figure 17: Grid and time-step convergence test for both Energy (top row)and Enstropybased (bottom row) POD

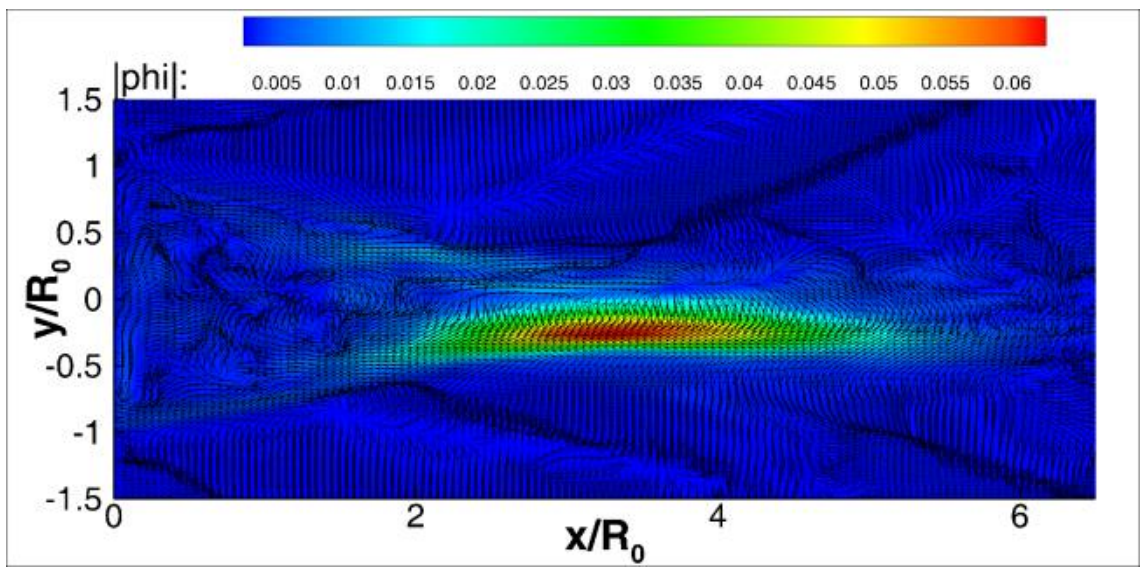

(a) 1st mode, Normalized Eigen value:0.068

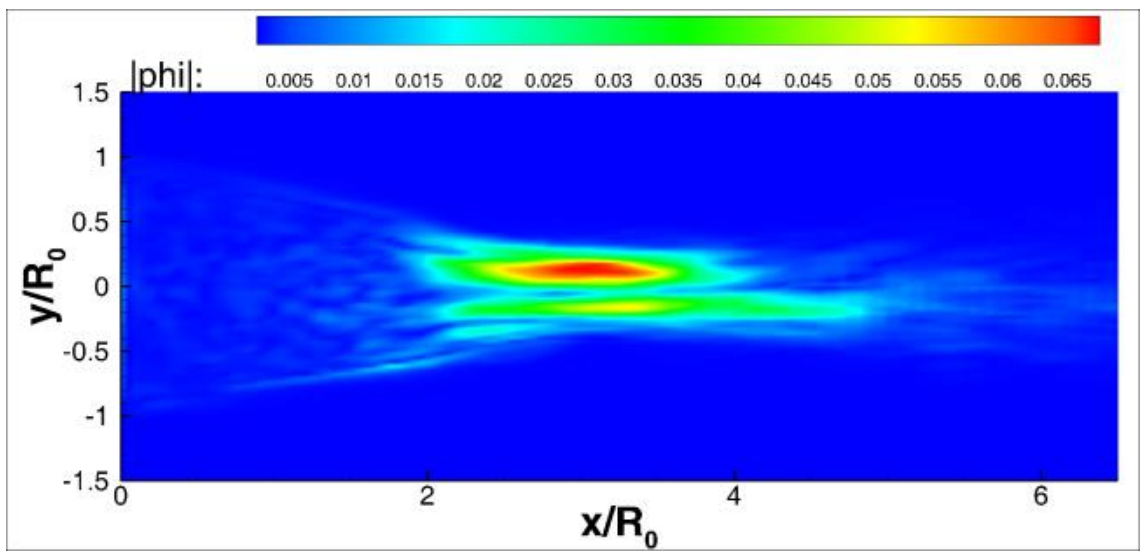

(a) 1st mode, Normalized Eigen value:0.045

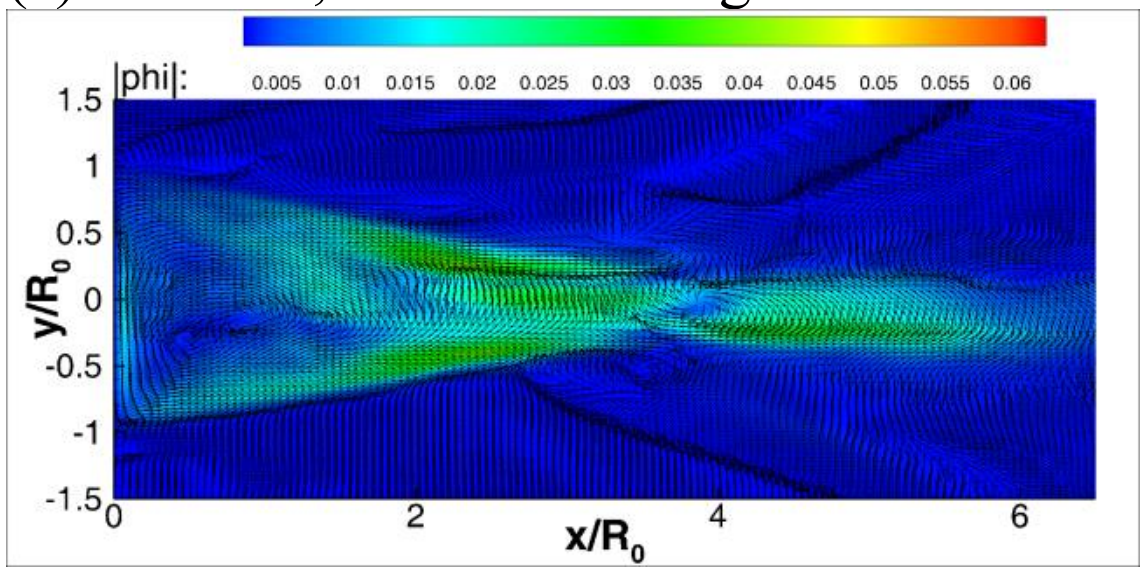

(b) 2nd mode, Normalized Eigen value:0.044

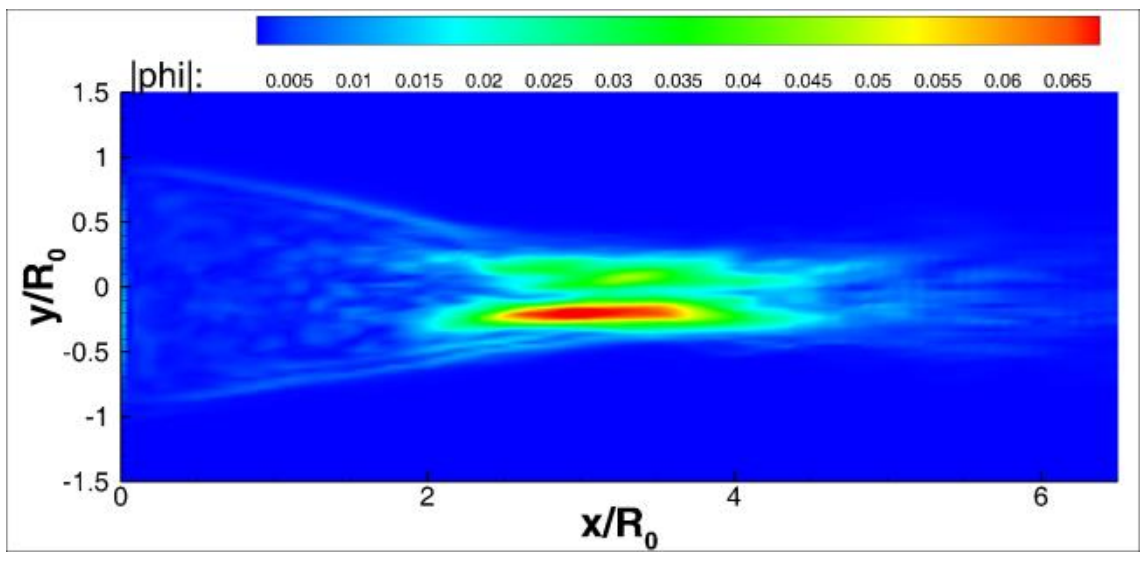

(b) 2nd mode, Normalized Eigen value:0.033

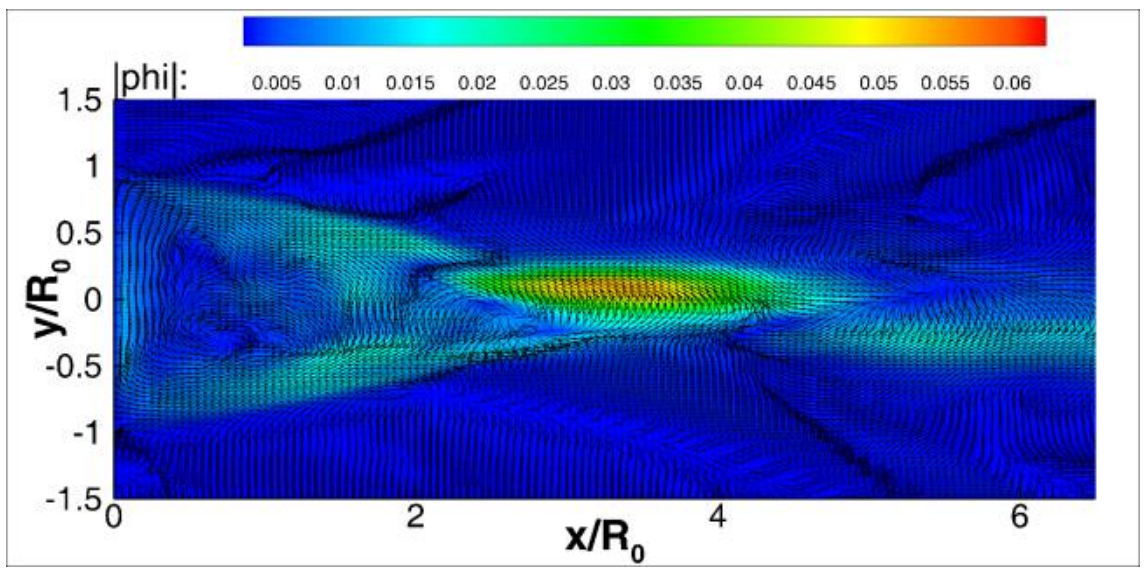

(c) 3rd mode, Normalized Eigen value:0.041

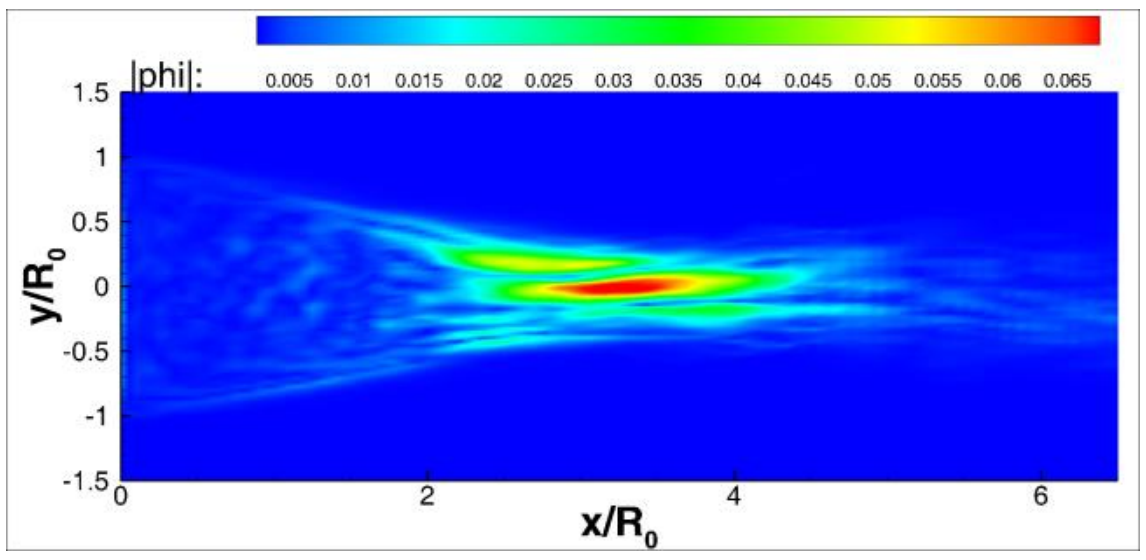

(c) 3rd mode, Normalized Eigen value:0.027

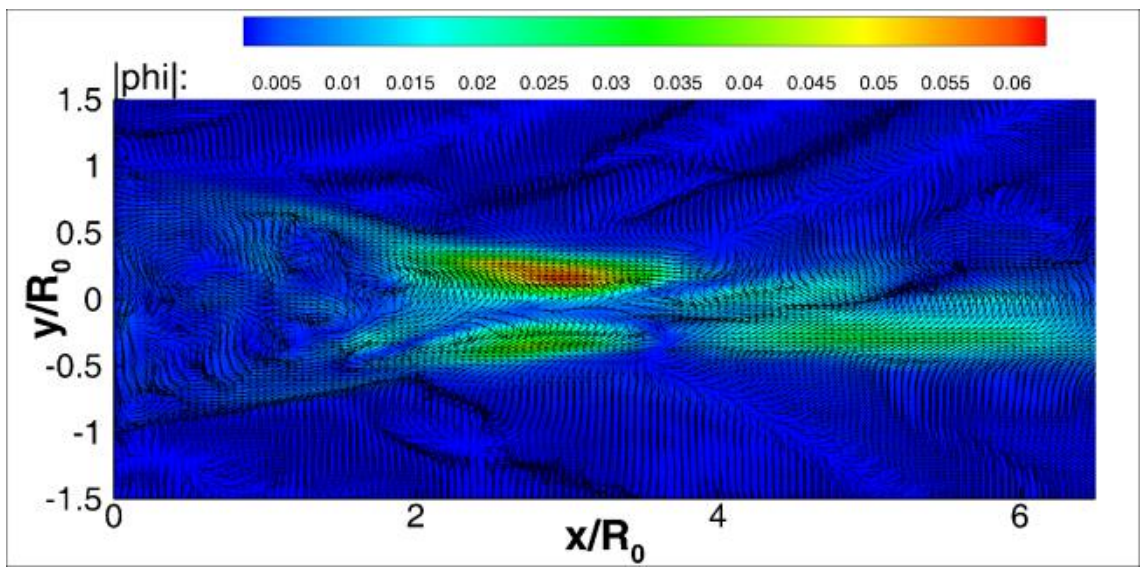

(d) 4th mode, Normalized Eigen value:0.032

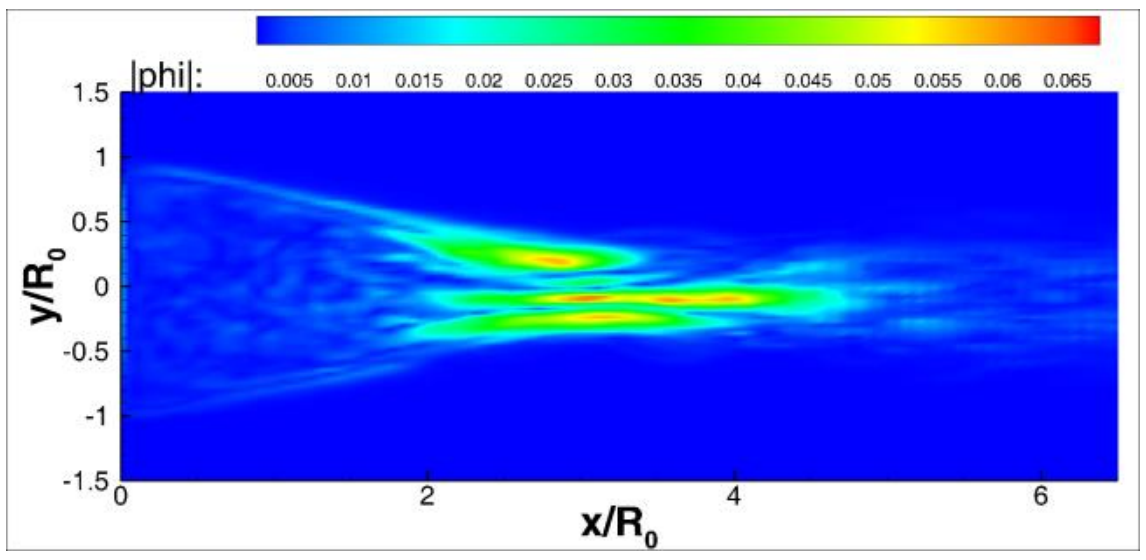

(d) 4th mode, Normalized Eigen value:0.023

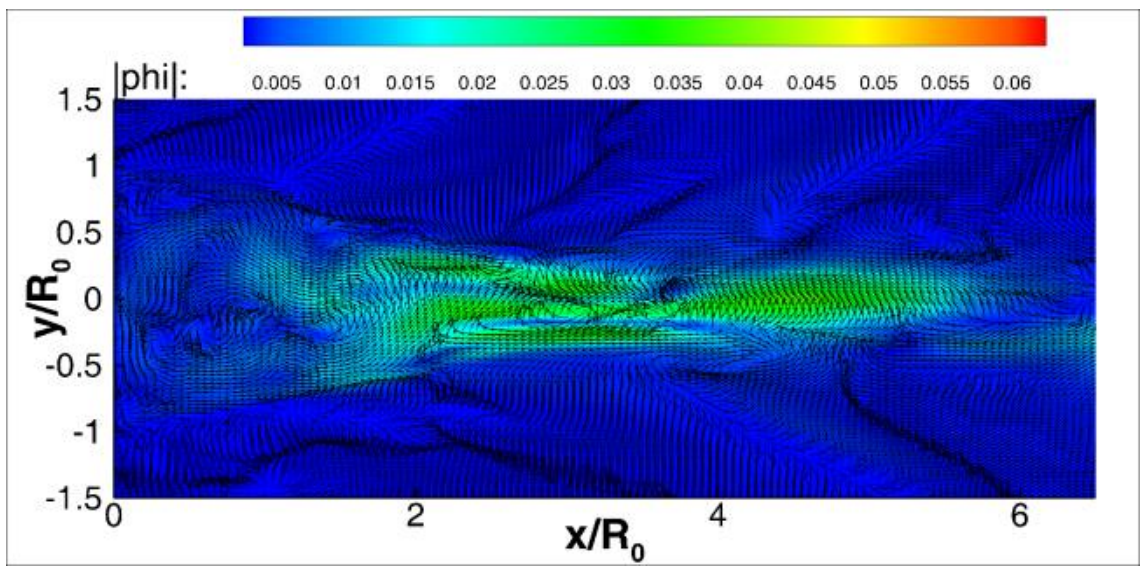

(e) 5th mode, Normalized Eigen value:0.029

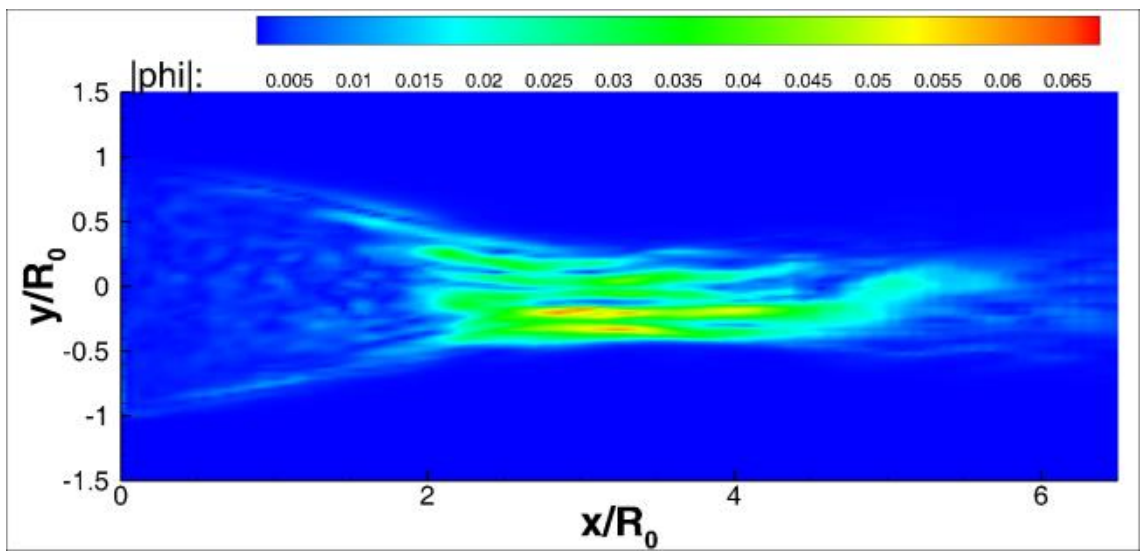

(e) 5th mode, Normalized Eigen value:0.019

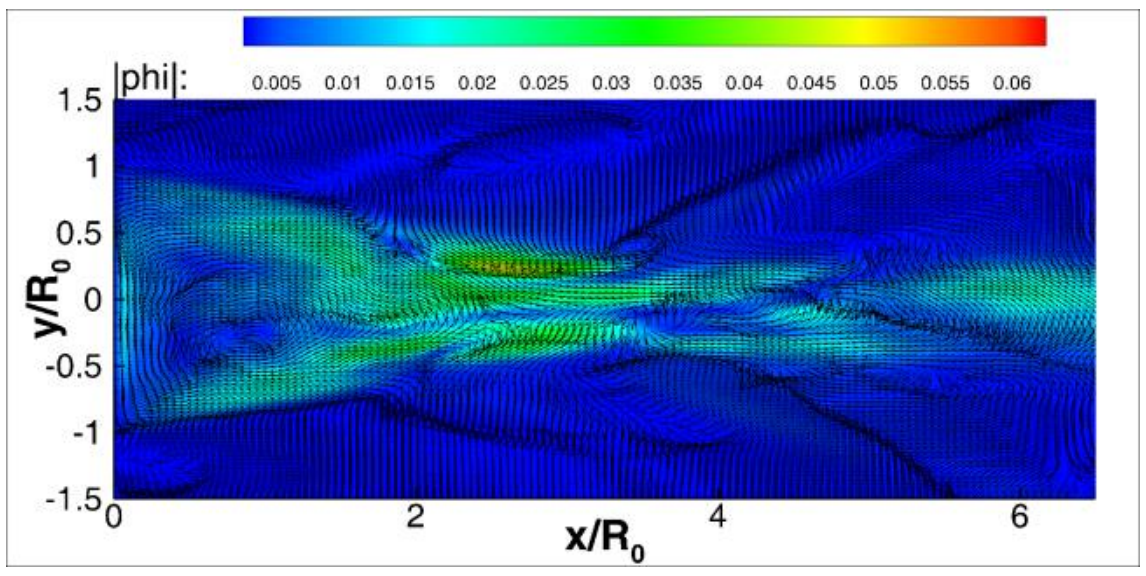

(f) 6th mode, Normalized Eigen value:0.024

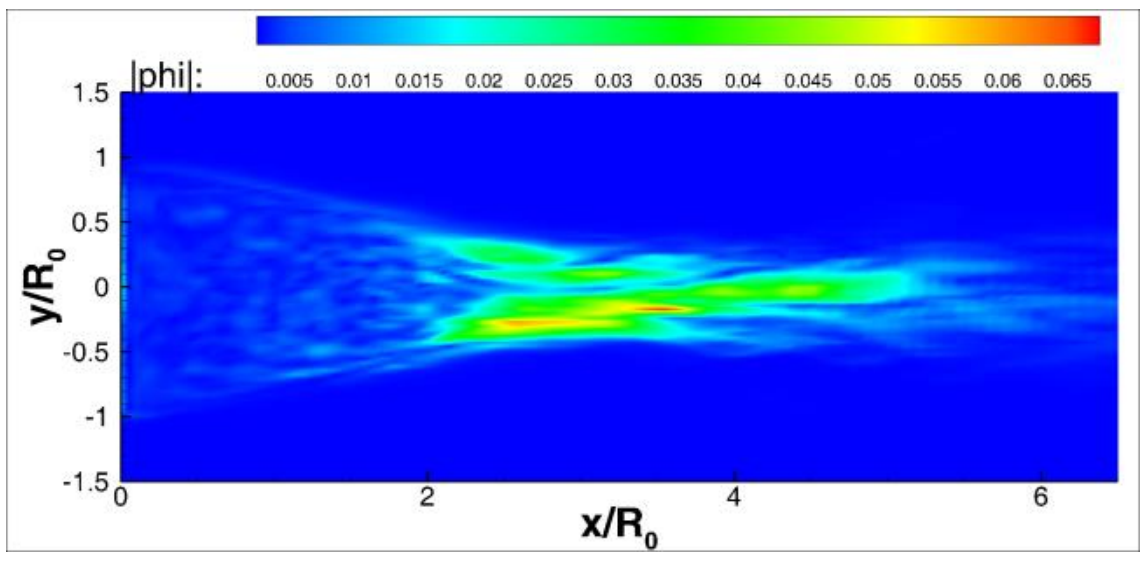

(f) 6th mode, Normalized Eigen value:0.018

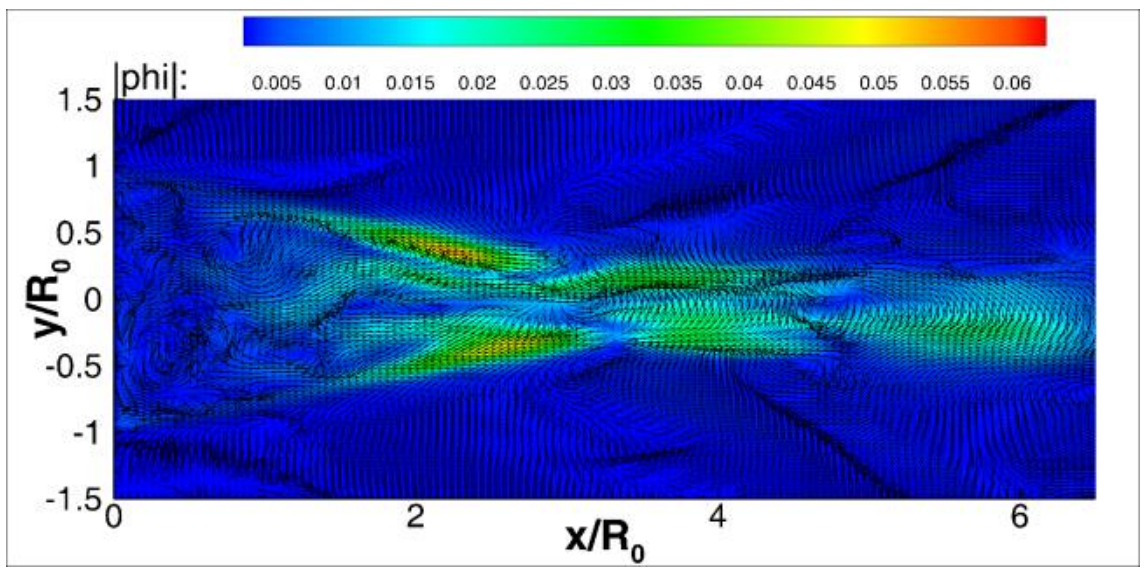

(g)7th mode, Normalized Eigen value:0.0221

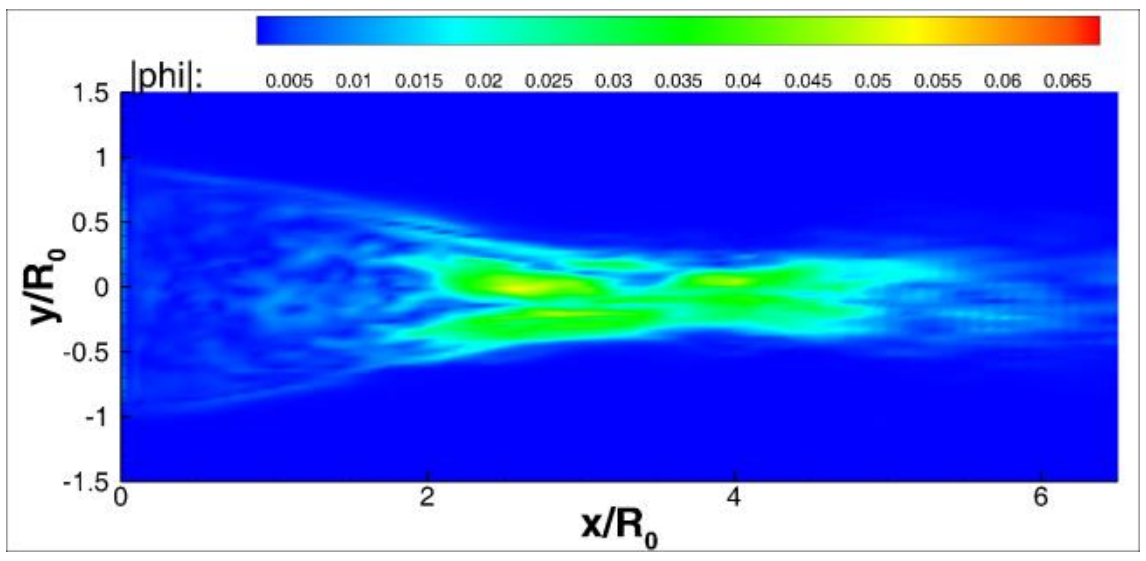

(g) 7th mode, Normalized Eigen value:0.016

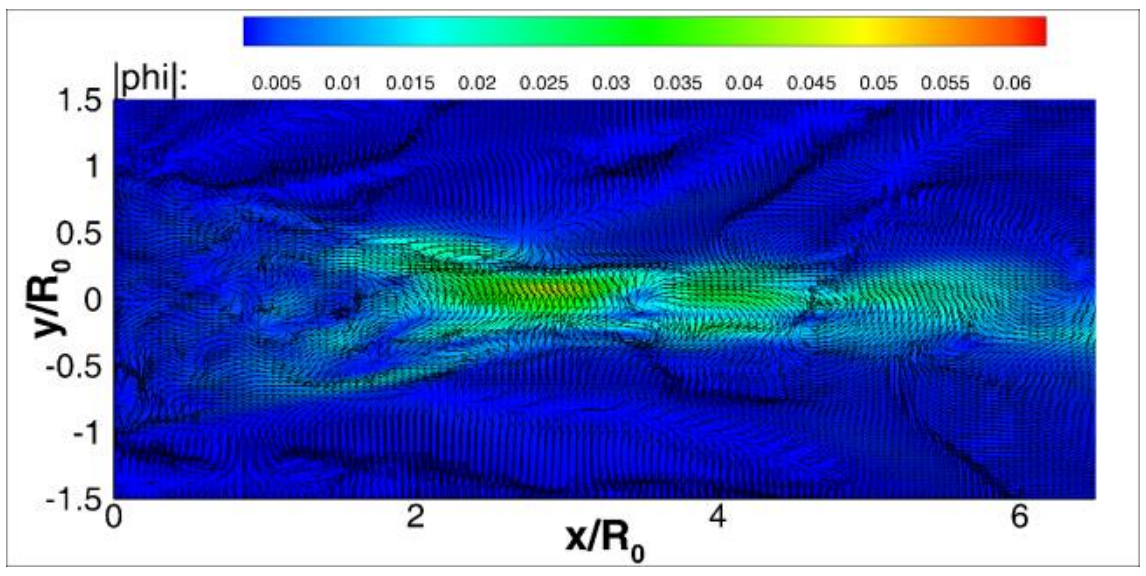

(h) 8th mode, Normalized Eigen value:0.021

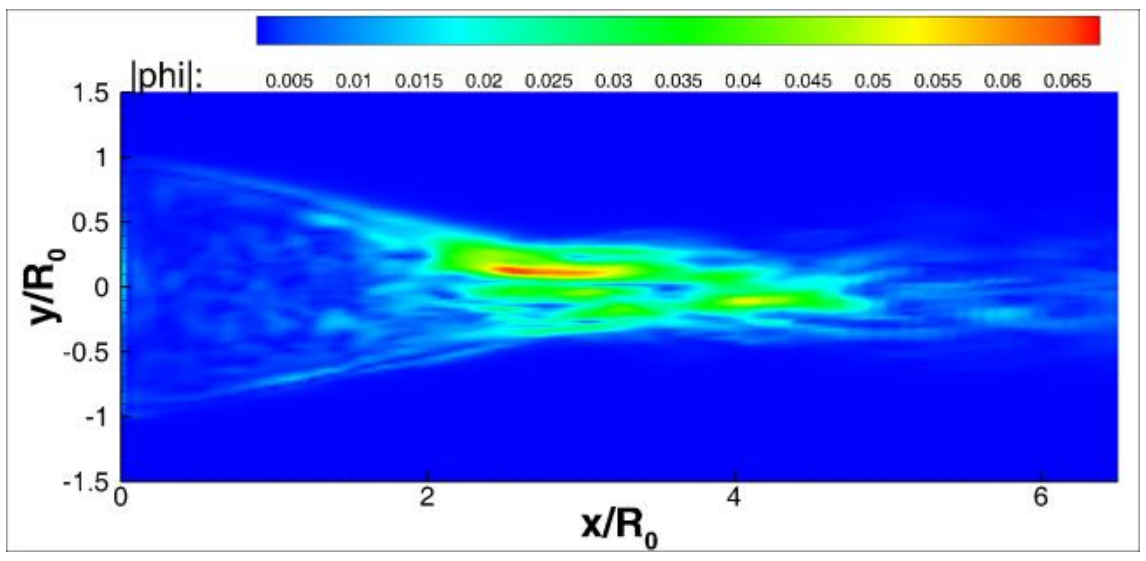

(h) 8th mode, Normalized Eigen value:0.0155

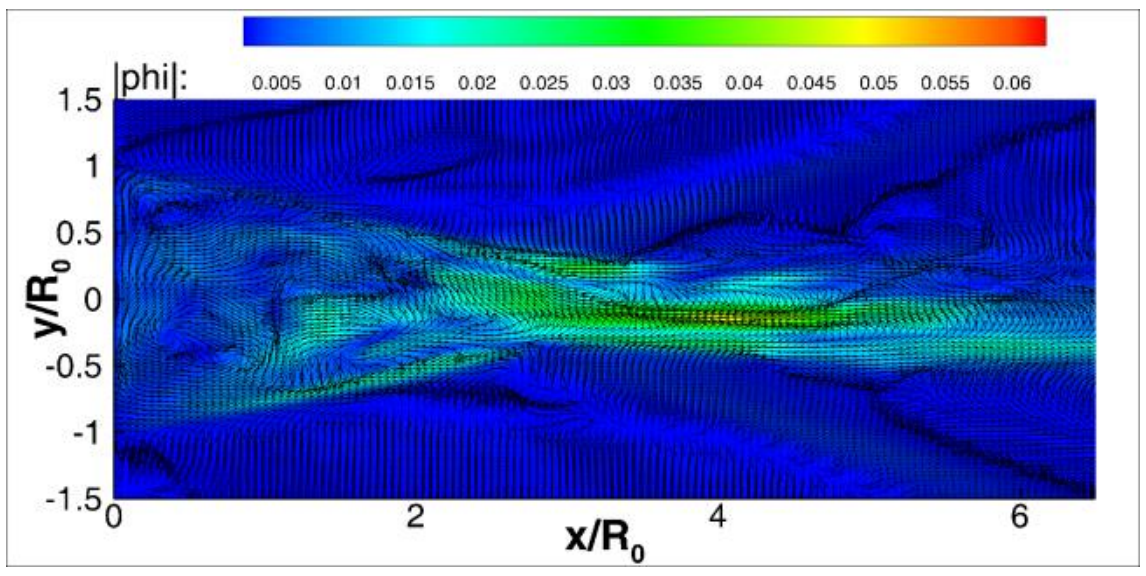

(i) 9th mode, Normalized Eigen value:0.019

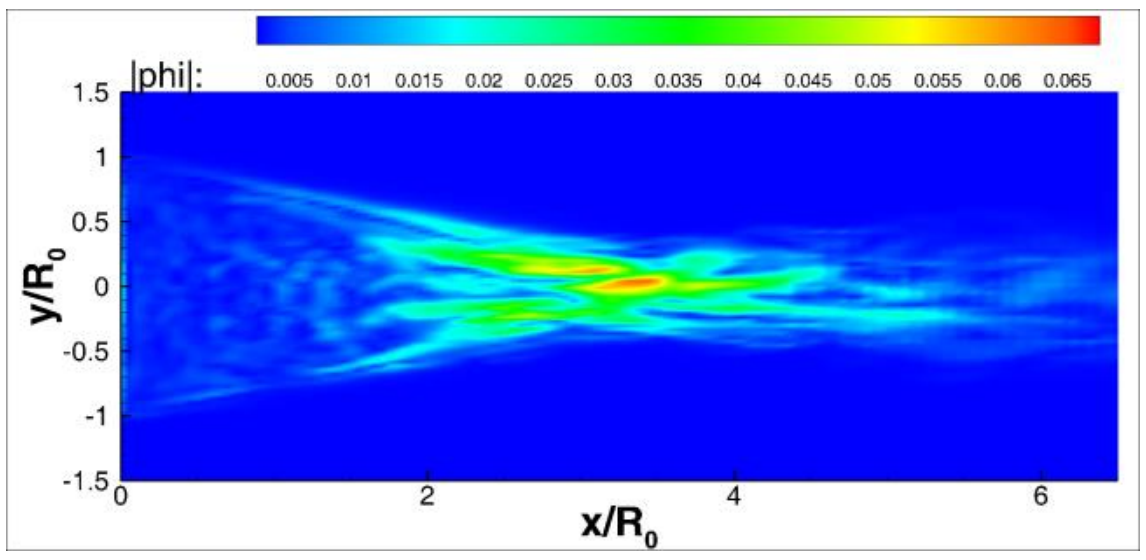

(i) 9th mode, Normalized Eigen value:0.014

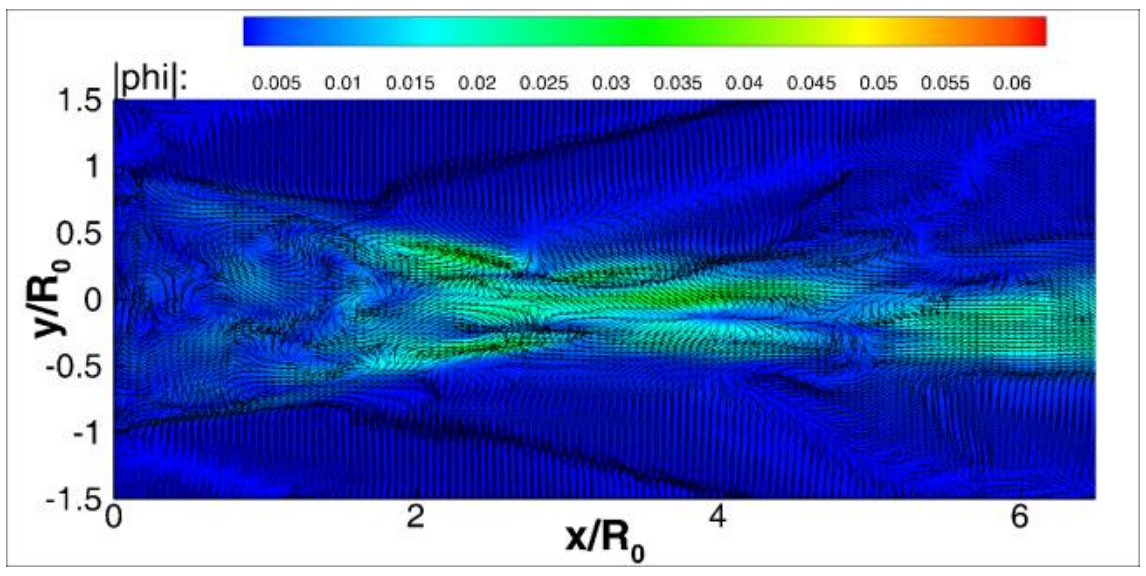

(j) 10th mode, Normalized Eigen value:0.018

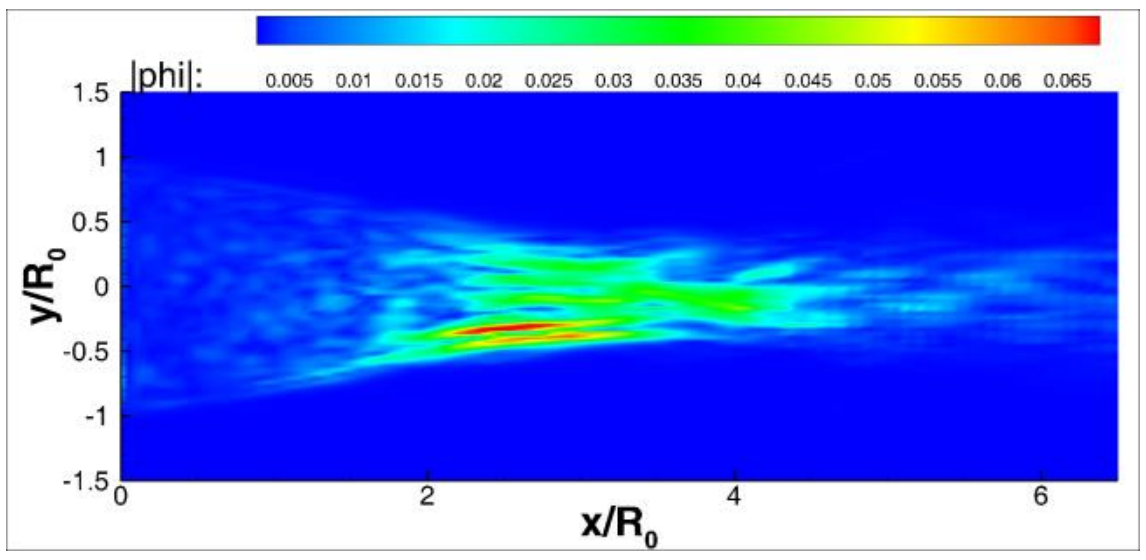

(j) 10th mode, Normalized Eigen value: 0.0139

Figure 18: Different modes obtained from both energy based (left column) and enstropy based (right column) POD